\documentclass[aps,twocolumn,superscriptaddress,prl, nofootbibin]{revtex4-1}
\usepackage{amssymb}
\usepackage{amsmath}
\usepackage{graphicx}
\usepackage{MnSymbol}
\usepackage{bm}
\usepackage[usenames]{color}

\begin{document}

\title{Spin-orbit interaction in InSb nanowires}
\author{I.~van Weperen}
\affiliation{QuTech and Kavli Institute of Nanoscience, Delft University of Technology, 2600 GA Delft, The Netherlands}
\author{B.~Tarasinski}
\affiliation{Instituut-Lorentz, Universiteit Leiden, P.O. Box 9506, 2300 RA Leiden, The Netherlands}
\author{D.~Eeltink}
\affiliation{QuTech and Kavli Institute of Nanoscience, Delft University of Technology, 2600 GA Delft, The Netherlands}
\author{V.~S.~Pribiag}
\altaffiliation{present address: School of Physics and Astronomy, University of Minnesota, 116 Church Street S.E., Minneapolis 55455, USA}
\affiliation{QuTech and Kavli Institute of Nanoscience, Delft University of Technology, 2600 GA Delft, The Netherlands}
\author{S.~R.~Plissard}
\altaffiliation{present address: Laboratoire d'Analyse et d'Architecture des Syst\`{e}mes, 7 Avenue du Colonel Roche, BP 54200 31031, Toulouse, France}
\affiliation{QuTech and Kavli Institute of Nanoscience, Delft University of Technology, 2600 GA Delft, The Netherlands}
{\affiliation{Department of Applied Physics, Eindhoven University of
Technology, 5600 MB Eindhoven, The Netherlands}
\author{E.~P.~A.~M.~Bakkers}
\affiliation{QuTech and Kavli Institute of Nanoscience, Delft University of Technology, 2600 GA Delft, The Netherlands}
\affiliation{Department of Applied Physics, Eindhoven University of
Technology, 5600 MB Eindhoven, The Netherlands}
\author{L.~P.~Kouwenhoven}
\affiliation{QuTech and Kavli Institute of Nanoscience, Delft University of Technology, 2600 GA Delft, The Netherlands}
\author{M.~Wimmer}
\email{m.t.wimmer@tudelft.nl}
\affiliation{QuTech and Kavli Institute of Nanoscience, Delft University of Technology, 2600 GA Delft, The Netherlands}

\date{\today}

\begin{abstract}
We use magnetoconductance measurements in dual-gated InSb nanowire
devices together with a theoretical analysis of weak antilocalization
to accurately extract spin-orbit strength. In particular, we show that magnetoconductance in our three-dimensional wires is very different compared to wires in two-dimensional electron gases. We obtain a large Rashba spin-orbit strength of $0.5 -1\,\text{eV\r{A}}$ corresponding to a spin-orbit energy of $0.25-1\,\text{meV}$. These values underline the potential of InSb nanowires in the study of Majorana fermions in hybrid semiconductor-superconductor devices.
 
\end{abstract}

\maketitle

Hybrid semiconductor nanowire-superconductor devices are a promising
platform for the study of topological superconductivity
\cite{Alicea2012}. Such devices can host Majorana fermions
\cite{Oreg2010,Lutchyn2010}, bound states with non-Abelian exchange
statistics. The realization of a stable topological state requires an
energy gap that exceeds the temperature at which experiments are
performed ($\sim$50 mK). The strength of the spin-orbit interaction
(SOI) is the main parameter that determines the size of this
topological gap \cite{Sau2012} and thus the potential of these devices
for the study of Majorana fermions. The identification of nanowire
devices with a strong SOI is therefore essential. This entails both
performing measurements on a suitable material and device geometry as
well as establishing theory to extract the SOI strength.

InSb nanowires are a natural candidate to create devices with a strong
SOI, since bulk InSb has a strong SOI
\cite{Winkler2003,Fabian2007}. Nanowires have been used in
several experiments that showed the first signatures of Majorana
fermions \cite{Mourik2012,Das2012,Deng2012,Churchill2013}.
Nanowires are either fabricated by etching out wires in planar
heterostructures or grown bottom-up. The strong confinement in the
growth direction makes etched wires two-dimensional (2D) even at high
density. SOI has been studied in 2D InSb wires \cite{Kallaher2010} and
in planar InSb heterostructures \cite{Kallaher2010b}, from which a SOI
due to structural inversion asymmetry \cite{Rashba1960}, a Rashba SOI
$\alpha_{R}$, of 0.03 eV\r{A} has been obtained
\cite{Kallaher2010b}. Bottom-up grown nanowires are three-dimensional
(3D) when the Fermi wavelength is smaller than the wire diameter. In
InSb wires of this type SOI has been studied by performing
spectroscopy on quantum dots \cite{Nilsson2009,Nadj-Perge2012}, giving
$\alpha_{R}$ = 0.16 -- 0.22 eV\r{A} \cite{Nadj-Perge2012}. However,
many (proposed) topological nanowires devices
\cite{Wimmer2011,Houzet2013,Hyart2013} contain extended conducting
regions, i.e. conductive regions along the nanowire much longer than
the nanowire diameter. The SOI strength in these extended regions has
not yet been determined. It is likely different from that in quantum
dots, as the difference in confinement between both geometries results
in a different effective electric field and thus different Rashba
SOI. Measurements of SOI strength in extended InSb nanowire regions
are therefore needed to evaluate their potential for topological
devices. Having chosen a nanowire material, further enhancement of
Rashba SOI strength can be realized by choosing a device geometry that
enhances the structural inversion asymmetry
\cite{Nitta1997,Engels1997}. Our approach is to use a
high-k dielectric in combination with a top gate that covers the InSb
nanowire.

\begin{figure}
\includegraphics[width=1.0\linewidth]{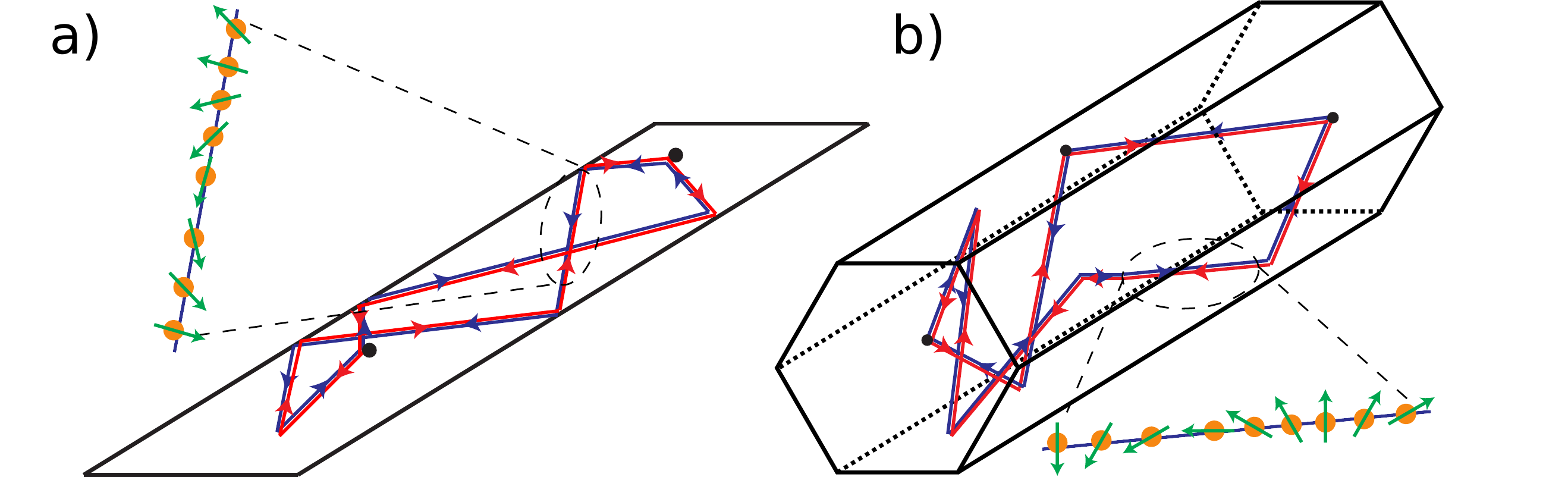}
\caption{Quantum interference along time-reversed paths in 2D (\textbf{a})) and 3D (\textbf{b)}) nanowires. In both cases an inversion symmetry induces spin precession in between (boundary) scattering events.}   
\label{fig1}
\end{figure}

The standard method to extract SOI strength in extended regions is
through low-field magnetoconductance (MC) measurements
\cite{Hikami1980,Iordanskii1994}. Quantum interference (see
Fig.~\ref{fig1}) in the presence of a strong SOI results in an
increased conductance, called weak anti-localization (WAL)
\cite{Bergmann1984}, that reduces to its classical value when a
magnetic field is applied \cite{Altshuler1981}. From fits of MC data
to theory a spin relaxation length is extracted. If spin relaxation
results from inversion asymmetry a spin precession length and SOI
strength can be defined. To extract SOI strength in nanowires the
theory should contain (1) the length over which the electron dephases
in the presence of a magnetic field, the magnetic dephasing length
\cite{Beenakker1988}, and (2) the relation between spin relaxation and
spin precession length \cite{Kettemann2007}. The magnetic dephasing
and spin relaxation length depend, besides magnetic field and SOI
strength respectively, on dimensionality and confinement. For
instance, in nanowires, the spin relaxation length increases when the
wire diameter is smaller than the spin precession length
\cite{Kiselev2000,Schapers2006,Kettemann2007}. Therefore the spin
relaxation length extracted from WAL is not a direct measure of SOI
strength. These effects have been studied in 2D wires
\cite{Beenakker1988,Kettemann2007}, but results for 3D wires are
lacking. As geometry and dimensionality are different (see
Fig.~\ref{fig1}), using 2D results for 3D wires is unreliable. Thus,
theory for 3D wires has to be developed.

In this Letter, we first theoretically study both magnetic dephasing
and spin relaxation due to Rashba SOI in 3D hexagonal nanowires. We
then use this theory to determine the spin-orbit strength from our
measurements of WAL in dual-gate InSb nanowire devices, finding a
strong Rashba SOI $\alpha_\mathrm{R}$ = $0.5 - 1\,\text{eV\r{A}}$.

The WAL correction to the classical conductivity can be computed
in the quasiclassical theory as \cite{Chakravarty86,
Beenakker1988, Kurdak92}
\begin{align}\label{eq:kurdakexpr}
\Delta &G = - \frac{e^2}{h} \frac{1}{L} \biggl[
3\, \biggl( \frac{1}{l_\varphi^2} + \frac{4}{3l_\text{so}^2} + \frac{1}{l_B^2} \biggr)^{-\frac{1}{2}}
\!\!\!- \biggl( \frac{1}{l_\varphi^2} + \frac{1}{l_B^2} \biggr)^{-\frac{1}{2}} \nonumber\\ 
&-3\, \biggl( \frac{1}{l_\varphi^2} + \frac{4}{3l_\text{so}^2} + \frac{d}{l_e^2} +  \frac{1}{l_B^2} \biggr)^{-\frac{1}{2}} 
\!\!\!+ \biggl( \frac{1}{l_\varphi^2} + \frac{d}{l_e^2} + \frac{1}{l_B^2} \biggr)^{-\frac{1}{2}}
\biggr]\,.
\end{align}
The length scales in this expression are the nanowire length $L$,
the mean free path $l_e$, the phase coherence length $l_\varphi$,
the magnetic dephasing length $l_B$, and the spin
relaxation length $l_\text{so}$. The mean free path $l_e = v_\text{F} \tau_e$
where $\tau_e$ is the mean time between scattering events and
$v_\text{F}$ the Fermi velocity. In addition, the remaining
length scales are also related to corresponding time scales
as
\begin{equation}\label{eq:diff_length}
l_{B,\varphi,\text{so}} = \sqrt{D\tau_{B,\varphi,\text{so}}}.
\end{equation}
where $D=\frac{1}{d} v_\text{F} l_e$ the diffusion constant
in $d$ dimensions ($d=3$ for bottom-up grown nanowires).


In the quasiclassical theory, $\tau_\varphi$ (and hence
$l_\varphi$) is a phenomenological parameter. In contrast, $\tau_B$
and $\tau_\text{so}$ are computed from a microscopic Hamiltonian,
by averaging the quantum mechanical propagator over classical
trajectories (a summary of the quasiclassical theory is given in the
supplemental material \cite{SI}). $\tau_B$ and $\tau_\text{so}$ thus
depend not only on microscopic parameters (magnetic field $B$ and
SOI strength, respectively), but through the average over
trajectories also on dimensionality, confinement, and $l_e$. We focus
on the case where Rashba SOI due to an effective electric field in the
$z$-direction, perpendicular to wire and substrate, dominates. Then
the microscopic SOI Hamiltonian is $\frac{\alpha_\mathrm{R}}{\hbar}
(p_x \sigma_y - p_y \sigma_x)$, where $\sigma_{x,y}$ are Pauli
matrices and $p_{x,y}$ the momentum operators. The corresponding
spin-orbit precession length, $l_\text{R} $, equals $\hbar^2/m^*
\alpha_\mathrm{R}$. In our treatment we neglect the Zeeman splitting,
$E_\text{Z}$ since we concentrate on the regime of large Fermi wave vector,
$k_\text{F}$, such that $\alpha_\mathrm{R} k_\text{F} \gg E_\text{Z}$.


The quasiclassical description is valid if the Fermi wave length
$\lambda_\text{F} \ll l_e, l_\text{R}$, and much smaller than
the transverse extent $W$ of the nanowire, i.e.~for many
occupied subbands. In particular, the quasiclassical method remains valid
even if $l_\text{R} <l_e,W$ \cite{Zaitsev05}.
Additional requirements are given in \cite{SI}.

We evaluate $\tau_B$ and $\tau_\text{so}$ numerically by averaging
over random classical paths for a given nanowire geometry.  The paths
consist of piece-wise linear segments of freely moving electrons with
constant speed \cite{Chakravarty86,Beenakker91}, only scattered
randomly from impurities and specularly at the boundary (for numerical
details see \cite{SI}). These assumptions imply a uniform electron
density in the nanowire.  Specular boundary reflection is expected as
our wires have no surface roughness \cite{Xu2012}.

\begin{figure}
\includegraphics[width=\linewidth]{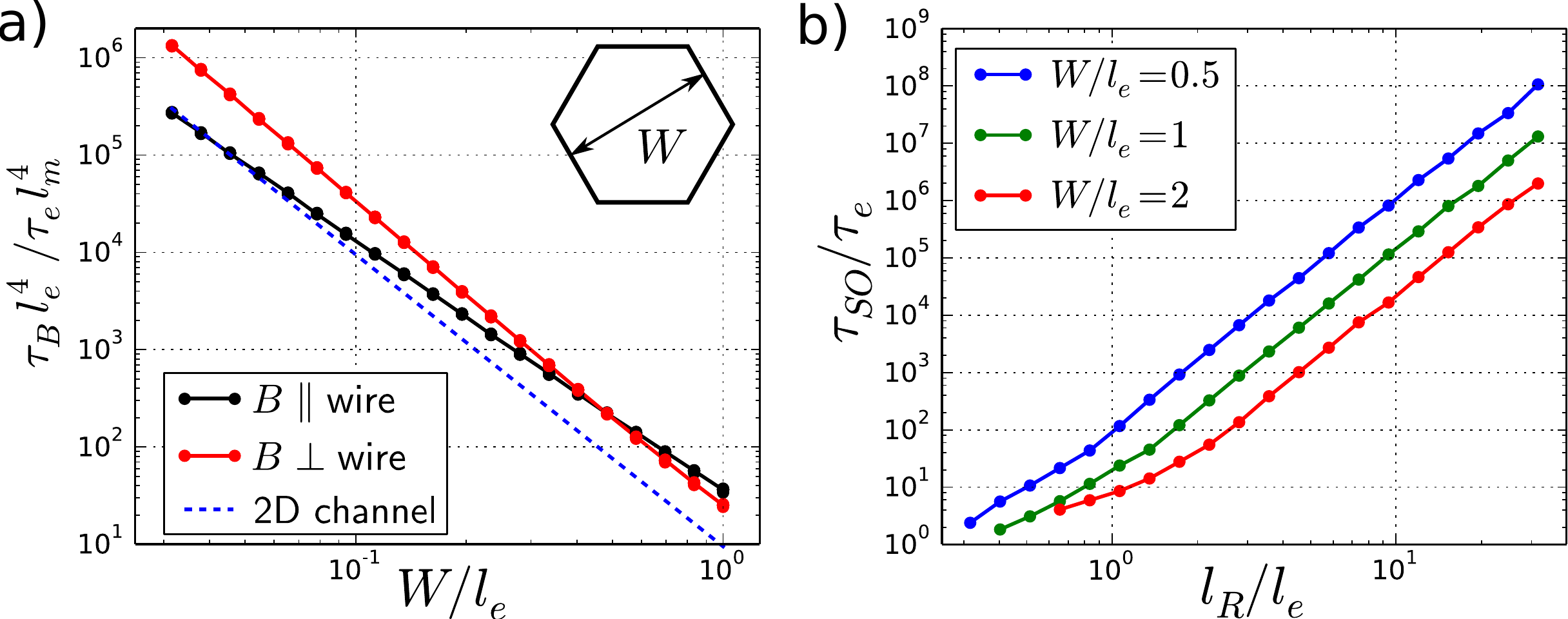}
\caption{a) Normalized dephasing time $\tau_Bl_e^4/\tau_e l_m^4$
  as a function of $W/l_e$ for a hexagonal nanowire (see inset) for field
  parallel (black) and perpendicular (red) to the nanowire.
  Dots are numerical data for different $l_m$ in the range $1-10^{2.5}$
  (10-20 points per $W$), solid lines a fit to
  Eq.~\eqref{eq:fit}. Dashed line is the 2D wire result of
  \cite{Beenakker1988}. b) $\tau_\text{so}/\tau_e$ as a function of
  spin-orbit strength $l_\mathrm{R}/l_e$ and different wire diameters in a
  3D hexagonal nanowire.
}\label{fig2}
\end{figure}


We apply our theory to nanowires with a hexagonal cross-section and
diameter $W$ (see inset in Fig.~\ref{fig2}(a)) in the
quasi-ballistic regime, $l_e \gtrsim W$. Fig.~\ref{fig2}(a) shows
the magnetic dephasing time $\tau_B$ (normalized by $\tau_el_m^4/l_e^4$
with $l_m =\sqrt{\hbar/eB}$) as a function of wire diameter. Both parallel and
perpendicular field give rise to magnetic dephasing due to the
three-dimensionality of the electron paths, in contrast to
two-dimensional systems where only a perpendicular field is relevant
(see Fig.~\ref{fig1}). The different field directions show
a different dependence on $W$, with, remarkably, $\tau_B$ (and thus
$l_B$) independent of field-orientation for
$W/l_e=0.5$. Our results for $\tau_\text{so}$ as a function of
$l_\text{R}$ are shown in Fig.~\ref{fig2}(b). We find an increase
of $\tau_\text{so}$ as the wire diameter $W$ is decreased, indicating
that confinement leads to increased spin relaxation times.


For $l_{m,\text{R}}, l_e \gtrsim W$ we can fit our results reliably as
\begin{equation}\label{eq:fit}
\tau_{B,\text{so}} = C \frac{l_{m,\text{R}}^4}{W^\gamma l_e^{(4-\gamma)}}\,.
\end{equation}
This is shown for $\tau_B$ in Fig.~\ref{fig2}(a)
where data for different $l_m$ and $W$ collapse to one line. In
particular for $\tau_B$, we find $C=34.1\pm 0.1$ and $\gamma=2.590\pm
0.002$ for parallel field, $C=22.3\pm 0.3$ and $\gamma=3.174\pm 0.003$
for perpendicular field. For $\tau_\text{so}$ $C=8.7 \pm 0.5$ and
$\gamma=3.2 \pm 0.1$. Note that our numerics is valid beyond the
range where the fit \eqref{eq:fit} is applicable. For example,
for $l_\text{R} \lesssim W$ the numerical result deviates from the
power-law of \eqref{eq:fit} as seen in Fig.~\ref{fig2}(b); in
this regime only the numerical result can be used.

The fit \eqref{eq:fit} allows for a quantitative comparison of our 3D
wire results to 2D wires: Both are similar in that there is flux
cancellation ($\gamma>2$) \cite{Beenakker1988} and suppressed spin relaxation due to
confinement. However, they exhibit a significantly different power-law.
As an example, in Fig.~\ref{fig2}(a) we compare to the 2D wire
result for weak fields from \cite{Beenakker1988} ($C=10.8$,
$\gamma=3$) that can differ by an order of magnitude from our
results. This emphasizes the need for an accurate description of
geometry for a quantitative analysis of WAL.

\begin{figure}
\includegraphics[width=\linewidth]{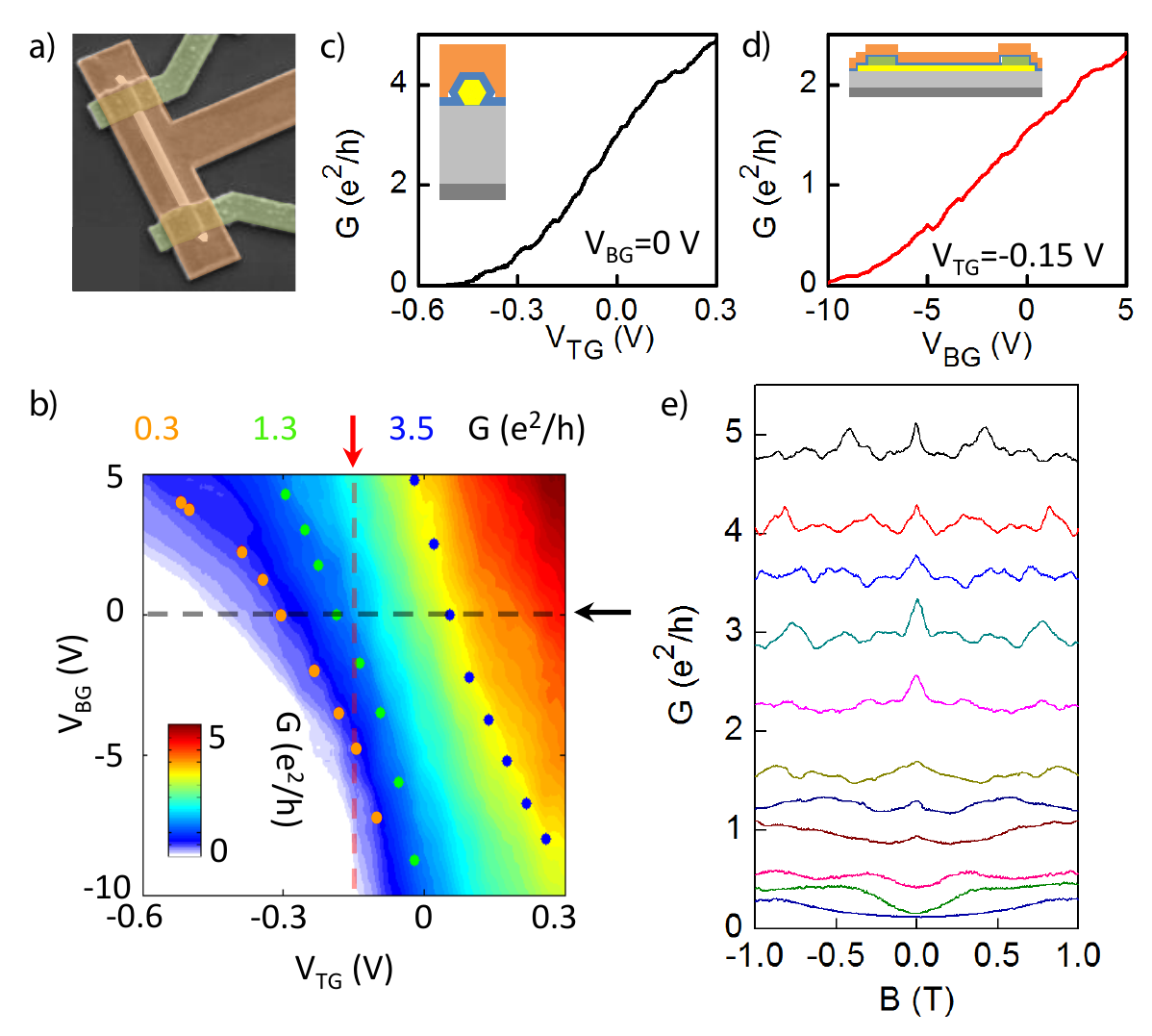}
\caption{(a) False color scanning electron microscopy image of
  device I. Contact spacing is $2\,\text{$\mu$m}$. Device fabrication is
  described in \cite{SI}. (b) Conductance $G$, as a function
  of top gate voltage, $V_{TG}$, and back gate voltage,
  $V_{BG}$. Arrows and dashed lines indicate cross sections shown in
  panels (c) and (d). Dots indicate voltages ($V_{BG}$,$V_{TG}$)
  at which traces in Fig.~\ref{fig4}(a) were taken (same dot color
  corresponds to same $G$). Data taken with 10 mV voltage bias at a
  temperature of $4.2\,\text{K}$. (c) $G$ as a function of $V_{TG}$ at
  $V_{BG} = 0\,\text{V}$. Inset: radial cross section of the device. The blue
  layer is HfO$_2$. (d) $G$ as a function of $V_{BG}$ at
  $V_{TG} = -0.15\,\text{V}$. Inset: axial cross section of the
  device. (e) Conductance, as a function of magnetic
  field at several values of device conductance controlled by
  $V_{TG}$, $V_{BG} = 0\,\text{V}$. Data taken with AC excitation $V_{AC} =
  100\,\text{$\mu$V$_{RMS}$}$.}
\label{fig3}
\end{figure}

We continue with the experiment. InSb nanowires \cite{Plissard2012} with diameter $W \approx 100\,\text{nm}$ are deposited onto a
substrate with a global back gate. A large ($\geq 2\,\text{$\mu$m}$) contact
separation ensures sufficient scattering between source
and drain. After contact deposition a HfO$_2$ dielectric layer is
deposited and the device is then covered by metal, creating an
$\Omega$-shaped top gate (Fig. \ref{fig3}a and insets of
Fig. \ref{fig3}c-d). Nanowire conductance is controlled with top
and back gate voltage, reaching a conductance up to $\sim 5e^2/h$
(Fig. \ref{fig3}b). The device design leads to a strong top gate
coupling (Fig. \ref{fig3}c), while back gate coupling is weaker
(Fig. \ref{fig3}d). From a field-effect mobility of $\sim
\text{11,000}\,\text{cm$^2$/Vs}$ a ratio of mean free path
to wire diameter $l_e/W = 1-2$ is estimated \cite{SI,Plissard2013}.

\begin{figure}[t]
\includegraphics[width=\linewidth]{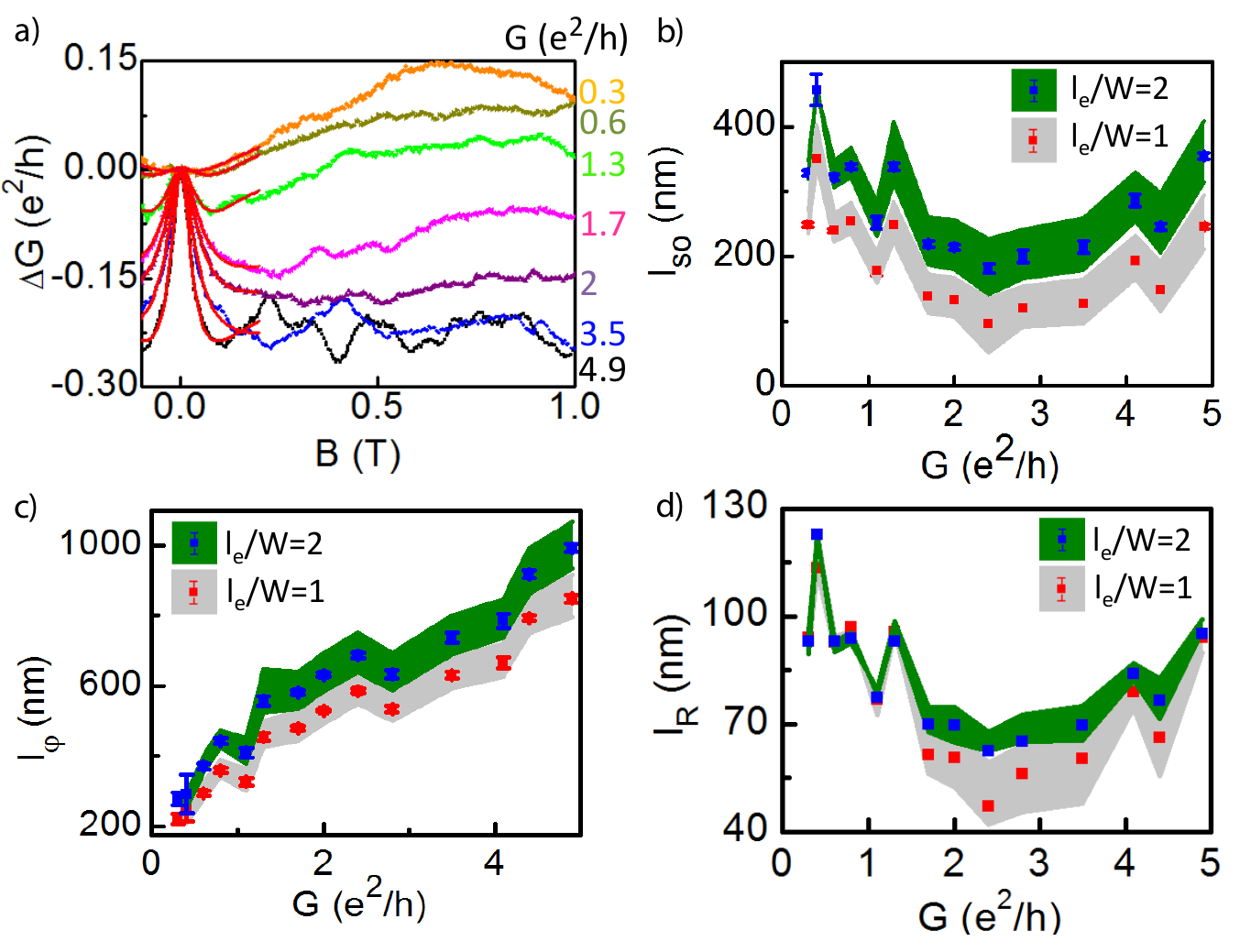}
\caption{(a) Magnetoconductance (MC) obtained after averaging
  MC traces taken at the same $G$. For $G$ = 3.5, 1.3 and $0.3e^2/h$
  the voltages at which these MC traces were taken are indicated in
  Fig. \ref{fig3}(b). Averaged MC traces have been centered to
  $\Delta G = 0$ at $B = 0\,\text{T}$. $G$($B=1\,\text{T}$) is indicated on the
  right. Red curves are fits to the data assuming $\frac{l_e}{W} = 1$.
  (b) Spin relaxation length $l_\mathrm{so}$ obtained from the fits of panel (a)
  ($\frac{l_e}{W} = 1$, blue points) and obtained from fits with
  $\frac{l_e}{W} = 2$ (red points). Standard deviation of the fit
  outcomes is indicated. The distribution around the blue and red
  points (green and gray bands, respectively) is given by the spin-orbit
  lengths obtained from fits with an effective width 15 nm smaller
  (resulting in longer $l_\text{so}$) or larger (resulting in shorter
  $l_\text{so}$) than the expected wire width $W = 90\,\text{nm}$.
  (c) Phase coherence length, $l_{\varphi}$ and (d) spin precession
  length $l_\text{R}$ as a function of device conductance.
  Figure formatting is as in panel (b).}
\label{fig4}
\end{figure}

At large $G$ the magnetoconductance, measured with conductance controlled by the top
gate at a temperature $T = 4.2\,\text{K}$ and with $B$ perpendicular to the
nanowire and substrate plane, shows an
increase of conductance of $\sim$ 0.2 to $\sim$ $0.3e^2/h$ around $B
= 0$ (Fig. \ref{fig3}(e)). $G$($B$) is, apart from reproducible conducantance fluctuations, flat at $B >$ 200 mT, which is further evidence of specular boundary scattering \cite{Beenakker91}. On reducing conductance below $\sim 1.5e^2/h$ WAL becomes less pronounced and a crossover to WL is seen.
 
Reproducible conductance fluctuations, most clearly seen at larger $B$
(Fig. \ref{fig3}(e)), affect the WAL peak shape. To suppress these
fluctuations several ($7 - 11$) MC traces are taken at the same device
conductance (see Fig. \ref{fig3}(b)). After averaging these traces
WAL remains while the conductance fluctuations are greatly suppressed
(Fig. \ref{fig4}(a)). Also here on reduction of conductance a
crossover from WAL to WL is seen. Very similar results are obtained
when averaging MC traces obtained as a function of top gate voltage
with $V_{BG} = 0\,\text{V}$ \cite{SI}. We expect that several ($\sim 10$)
subbands are occupied at device conductance $G \gtrsim 2e^2/h$
(see \cite{SI}). Hence, our quasiclassical approach is valid
and we fit the averaged MC traces to
Eq.~\eqref{eq:kurdakexpr} with $l_\mathrm{so}$, $l_{\varphi}$ and the conductance at large magnetic
field $\Delta G(B\rightarrow\infty)$ as fit parameters. $l_{\mathrm{B}}$ is extracted from Eq.~\eqref{eq:fit}. Wire diameter and mean free path are fixed in each fit, but we extract fit results for a wire diameter deviating from its expected value and for both $\frac{l_e}{W} = 1$ and $\frac{l_e}{W} =
2$. We find good agreement between data and fits (see
Fig. \ref{fig4}(a)). While showing fit results covering the full range of $G$, we base our conclusions on results obtained in the quasiclassical transport regime $G \gtrsim$ 2e$^2$/h.

On increasing conductance, the spin relaxation length first decreases to
$l_\mathrm{so} \approx 100 - 200\,\text{nm}$, then increases again to
$l_\mathrm{so} \approx 200-400\,\text{nm}$ when $G\geq 2.5e^2/h$
(Fig.~\ref{fig4}(b)). The phase coherence length
(Fig.~\ref{fig4}(c)) shows a monotonous increase with device
conductance. This increase can be explained by the density dependence
of either the diffusion constant or the
electron-electron interaction strength \cite{Lin2002}, often reported as the
dominant source of dephasing in nanowires
\cite{Liang2012,Kallaher2010}.

Spin relaxation \cite{Wu2010} in our device can possibly occur via the
Elliot-Yafet \cite{ElliotYafet1963} or the D'yakonov-Perel'
mechanism \cite{Dyakanov1972}, corresponding to spin randomization at or in between
scattering events, respectively. The Elliot-Yafet contribution can be
estimated as $l_\mathrm{so,EY} =
\sqrt{\frac{3}{8}}\frac{E_{G}}{E_{F}}l_{e}\frac{(E_G+\Delta_\mathrm{SO})(3E_G+2\Delta_\mathrm{SO})}{\Delta_\mathrm{SO}(2E_G+\Delta_\mathrm{SO})}
\geq 300 - 600\,\text{nm}$ \cite{Chazalviel1975}, with band gap $E_G =
0.24\,\text{eV}$, Fermi energy $E_F \leq 100\,\text{meV}$, spin-orbit
gap $\Delta_\mathrm{SO} =0.8\,\text{eV}$ and $\frac{l_e}{W} = 1 - 2$. For the
D'yakonov-Perel' mechanism, we note that our nanowires have a
zinc-blende crystal structure, grown in the [111]
direction, where Dresselhaus SOI is absent for momentum along the
nanowire \cite{footnote_dresselhaus}.
We therefore expect that Rashba SOI is the dominant source of spin
relaxation, in agreement with previous
experiments \cite{Nadj-Perge2012}.  As found in our theoretical
analysis, it is then crucial to capture confinement
effects accurately. Our $l_\mathrm{so}$ correspond to $\frac{\tau_\mathrm{so}}{\tau_e} =$ 2
-- 15 that are captured well by our simulations \cite{footnote1}. Given that $W \approx l_{\mathrm{R}}$, we extract the $l_R$ corresponding to our $\frac{\tau_\mathrm{so}}{\tau_e}$ directly from Fig. \ref{fig2}(b). We
extract spin precession lengths $l_\mathrm{R}$ of $50- 100\,\text{nm}$, shown
in Fig. \ref{fig4}(d), corresponding to $\alpha_\mathrm{R} = 0.5 -
1.0\,\text{eV\r{A}}$. MC measurements on a second device show very
similar $l_\mathrm{R}$ \cite{SI}.

\begin{figure}
\includegraphics[width=\linewidth]{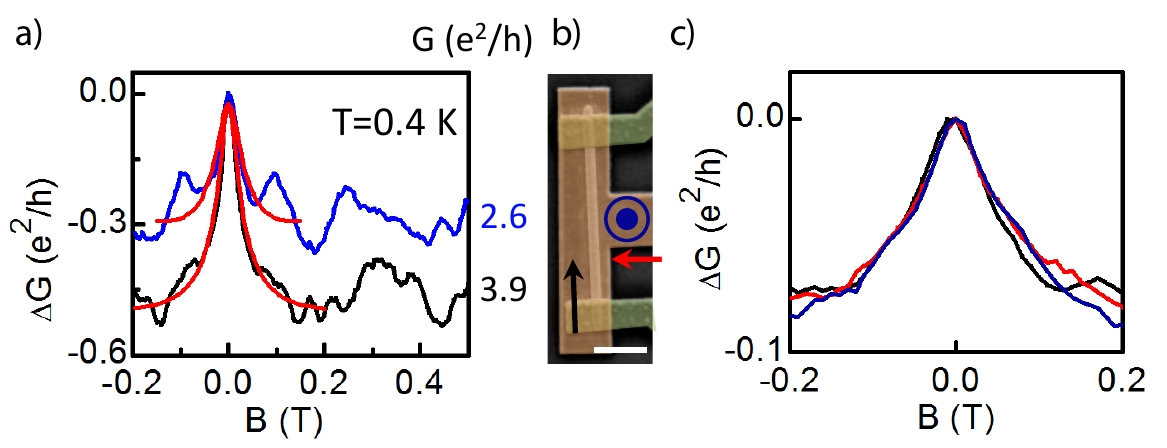}
\caption{(a) Magnetoconductance (MC) at $T = 0.4\,\text{K}$. Each MC
  trace is obtained after averaging 21 MC traces taken along the
  top-gate controlled pinch-off trace shown in Fig. \ref{fig3}(c)
  ($V_{BG} = 0\,\text{V}$). Black (blue) trace is the average of
  traces taken between $V_{TG} = 0.34\,\text{V}$ and $V_{TG} =
  0.14\,\text{V}$ ($V_{TG} = 0.12\,\text{V}$ and $V_{TG} =
  -0.08\,\text{V}$) with steps of $20\,\text{mV}$. The voltage excitation
  $V_{AC}$ was $10\,\mu$\text{V}$_{RMS}$. $G$($B = 0.5\,\text{T}$) is
  indicated on the right. Phase coherence and spin relaxation length
  obtained from fits (in red) to the traces is $1078 \pm 32$ ($1174
  \pm 39$) nm and of $95 \pm 18$ ($205 \pm 16$) nm respectively for
  $\frac{l_e}{W} = 1$ (2). Values obtained at $G = 2.6e^2/h$ are given
  in \cite{SI}. (b) False color scanning electron microscope image of
  device II with different magnetic field orientations indicated by
  the arrows. Scale bar is $1\,\mu$m.(c) MC obtained with $B$ parallel
  to the nanowire (in-plane angle w.r.t. nanowire $\theta \approx
  5^{\circ}$, black), $B$ perpendicular to the nanowire in the plane
  of the substrate ($\theta \approx 95^{\circ}$, red) and $B$
  perpendicular to the substrate plane (blue). $V_{TG}= 0.2\,\text{V}$,
  $V_{BG} = 0\,\text{V}$. Smaller $\Delta G$ compared to the preceding data is
  due to a larger contact resistance ($\sim 10\,\text{k$\Omega$}$) of this
  device for which no correction was made.}
	\label{fig5}
\end{figure}

To confirm the interpretation of our MC measurements we extract MC at
a lower temperature $T = 0.4\,\text{K}$ (Fig. \ref{fig5}a). We find larger
WAL amplitudes of up to $\Delta G\sim 0.5e^2/h$, while the width of
the WAL peak remains approximately the same as at $T = 4.2\,\text{K}$,
corresponding to a longer $l_{\varphi}$ at lower temperature, with
approximately constant $l_\mathrm{so}$. A longer $l_{\varphi}$ is expected at lower
temperature, as the rate of inelastic scattering,
responsible for loss of phase coherence, is reduced in this regime.

Our theoretical analysis found similar dephasing times for magnetic fields perpendicular and
parallel to the nanowire for our estimated mean free paths, $l_e/W
= 1 - 2 $. Indeed, we observe virtually identical WAL for fields
parallel and perpendicular to the nanowire in our second device (see
Figs.~\ref{fig5}(b)-(c)). WAL in the first device is also very similar for both field directions \cite{SI}. This is in striking contrast to MC measurements in
two-dimensional systems where only a perpendicular magnetic field
gives strong dephasing due to orbital effects. It also provides strong
support for the assumptions made in our theory, and emphasizes the
importance of including the three-dimensional nature of nanowires to
understand their MC properties. In contrast, WL is anisotropic \cite{SI}, which we attribute to a
different density distribution at low conductance compared to the high
conductance at which WAL is seen.

Relevant to Majorana fermion experiments
is the spin-orbit energy, $E_\mathrm{SO} = \frac{m\alpha_\mathrm{R}^2}{2\hbar^2}$,
that is $0.25 - 1\,\text{meV}$ in our devices. These values compare
favorably to InAs nanowires that yield $\alpha_\mathrm{R}^{InAs} = 0.1 -
0.3\,\text{eV\r{A}}$ \cite{Liang2012,Hansen2005} and
corresponding $E_\mathrm{SO}^{InAs} = 15 -
135\,\mu\text{eV}$. $E_\mathrm{SO}^{InSb}$ is similar or slightly larger than
reported spin-orbit energies in Ge/Si core-shell nanowires
($E_\mathrm{SO}^{Ge/Si} = 90 - 600\,\mu\text{eV}$ \cite{Hao2010}), while
$\alpha_\mathrm{R}^{InSb}$ is larger than $\alpha_\mathrm{R}^{Ge/Si} = 0.07 -
0.18\,\text{eV\r{A}}$). Note that the device geometries and
expressions for $\alpha_\mathrm{R}$($l_\mathrm{so}$) used by different authors vary
and that often only $l_\mathrm{so}$, not $l_\mathrm{R}$ is evaluated. With our $E_\mathrm{SO}$ we then find, following the analysis of Ref.~\cite{Sau2012},  a topological gap of $\sim 0.1 - 1 \text{K}$ \cite{SI} even for our moderate mobilities of order
$10000\,\text{cm$^2$/Vs}$. This gap largely exceeds the temperature and previous estimates. Hence, our findings underline the potential of InSb
nanowires in the study of Majorana fermions.

We thank C.~M.~Marcus, P.~Wenk, K.~Richter and I.~Adagideli for discussions. Financial support for this work is provided by the Dutch Organisation for Scientific Research (NWO), the Foundation for Fundamental Research on Matter (FOM) and Microsoft Corporation Station Q. V.~S.~P. acknowledges funding from NWO through a Veni grant.

\onecolumngrid
\appendix

\setcounter{figure}{0}
\renewcommand\thefigure{S\arabic{figure}}
\newpage

\section{Supplemental material}

\section{1. Summary of the quasiclassical theory}

Within the quasiclassical formalism, the weak (anti)localization correction
$\Delta G$ is given as
\cite{suppChakravarty86, suppBeenakker91, suppKurdak92}
\begin{equation}\label{eq:quasiclassics}
\Delta G = - \frac{2 e^2}{\pi \hbar} \frac{D}{L} \int_0^\infty dt\,
C(t)\,(1-e^{-t/\tau_e})\, e^{-t/\tau_\varphi}
\llangle\mathcal{M}_{B}(t) \rrangle \,
\llangle\mathcal{M}_{\text{so}}(t) \rrangle
\end{equation}
In this expression, $L$ is the length of the nanowire,
$C(t) = (4\pi D t)^{-1/2}$ is the 1D return
probability, $D=\frac{1}{d} v_\text{F} l_e$ the diffusion
coefficient ($d=3$ for the nanowires). $\llangle \dots \rrangle$
denotes an average over all classical paths that close after time $t$. $\mathcal{M}_B$ is due to
the orbital effect of the magnetic field and reads \cite{suppChakravarty86}
\begin{equation}\label{eq:modB}
\mathcal{M}_B(t) = e^{i \phi(t)}\text{, with }
\phi(t) = \frac{2 e}{\hbar} \int_{\bm{x}(0)}^{\bm{x}(t)} \bm{A} \cdot d\bm{l}\,.
\end{equation}
The Hamiltonian of spin-orbit interaction (SOI) can in general be written
as
\begin{equation}\label{eq:soiham}
H_\text{SOI}= \bm{\sigma} \cdot \bm{B}_\text{so}(\bm{p})
\end{equation}
where $\bm{\sigma}$ is a vector of Pauli matrices and $\bm{B}_\text{so}$ a
momentum-dependent effective magnetic field due to the SOI. In the case
of Rashba SOI as considered here we have $\bm{B}_\text{so}(\bm{p}) =
\frac{\alpha_\mathrm{R}}{\hbar} (-p_y, p_x, 0)$.
The SOI of Eq.~\eqref{eq:soiham} then gives rise to the modulation
factor \cite{suppChakravarty86, suppZaitsev05}
\begin{align}\label{eq:modSOI}
&\mathcal{M}_\text{so}(t) =  \frac{1}{2} \text{Tr}\left(W(t)^2\right)\nonumber\\
&W(t) =\mathcal{T}  \exp{\left[\frac{i}{\hbar} \int_0^t dt' \bm{\sigma} \cdot
\bm{B}_\text{so}(\bm{p}(t))\right]}
\end{align}
where $\mathcal{T}$ is the time-order operator.

When the motion along the longitudinal direction of wire is diffusive,
the modulation factors generally decay exponentially with
time \cite{suppChakravarty86},
\begin{equation}\label{eq:expform}
\llangle \mathcal{M}_B(t) \rrangle = e^{-t/\tau_B}\text{, and }
\llangle \mathcal{M}_\text{so}(t) \rrangle = \tfrac{3}{2}\,e^{-4t/3\tau_\text{so}}
-\tfrac{1}{2}\,.
\end{equation}
Note that $\tau_B$ and $\tau_\text{so}$ depend explicitly on the
magnetic field $B$ and the SOI strength through equations \eqref{eq:modB}
and \eqref{eq:modSOI}, respectively. However, through the average
over classical paths, $\llangle \dots \rrangle$ they also depend on the
geometry of the nanowire and the mean free path $l_e$.

With the exponential form of the modulation factors in Eq.~\eqref{eq:expform}
the integral in Eq.~\eqref{eq:quasiclassics} can be performed to
give the expression (1) of the conductance correction in the main text.

\subsection{Requirements of the quasi-classical theory}

The quasiclassical description is valid if the Fermi wave length
$\lambda_\text{F}$ is much smaller than the typical transverse extent
of the nanowire $W$, i.e.~for many occupied subbands. It also requires
that the classical paths are neither affected by magnetic field nor
SOI: The former requires that the cyclotron radius $\lambda_\text{cyc}
>> W, l_e$ \cite{suppChakravarty86, suppBeenakker91}, the latter that the
kinetic energy dominates over the spin-orbit energy so that
$l_\text{R} \gg \lambda_F$ \cite{suppZaitsev05}.  In particular, the
quasiclassical method is valid also for $l_\text{R} <l_e,W$.
Additional requirements are $\tau_B, \tau_\text{so} \gg \tau_e$, for
the exponential decay of magnetic dephasing time (length) and spin
relaxation time to be valid \cite{suppBeenakker91, suppZaitsev05}. In
addition we must have $l_\varphi \gg W$ to be in the
quasi-one-dimensional limit, where the return probability $C(t)$ in
Eq.~\eqref{eq:quasiclassics} is given by the 1D return probabilty.

These are the fundamental requirements for the quasiclassical theory to
hold. They should not confused with the stronger requirements
$l_{m,\text{R},e} \gtrsim W$ needed for the validity of the fit in
Eq.~(3) of the main text.

\subsubsection{Experimental fulfilment of quasi-classical requirements}

The number of occupied subbands is discussed in section 4 of this
document. As shown in Fig.~4c of the main text, $l_\varphi$ largely
exceeds the wire diameter for a large range of conductance, thereby
obeying the requirement for a one-dimensional quantum interference
model. The range of $B$ (up to $200\,\text{mT}$) in the fits in
Figs.~4-5 of the main text and in the figures in this document in
general obey $\tau_B \gtrsim \tau_e$. Alternatively, fitting over a
smaller $B$-range (up to $75 - 100\,\text{mT}$, fulfilling $l_m\gtrsim W$,
$\tau_e$ and $\lambda_\text{cyc} >> W,l_e$ to a larger
extent) can be performed on MC traces showing WAL without WL at larger
$B$ (observed when $G\geq 2e^2/h$) with fixed $\Delta
G(B\rightarrow\infty)$, yielding the same results within $\sim$ 20\%.

\section{2. Monte Carlo evaluation of the weak (anti)localization correction.}

In order to obtain the decay times in Eq.~\eqref{eq:expform} as a
function of mean free path $l_e$, wire diameter $W$, and magnetic
field $B$ or Rashba spin-orbit strength $\alpha_R$, we performed
Monte-Carlo simulations of quasiclassical paths in a hexagonal
nano-wire, as has been described before in Refs.~\cite{suppChakravarty86,
suppBeenakker91, suppZaitsev05}.

\subsection{Model and Boltzmannian ensemble}
We model the nanowire as a three-dimensional prisma of infinite
length, with a regular hexagon as cross-section.

A Boltzmannian ensemble of quasiclassical paths is created, with each
path consisting of propagation along a sequence of straight line
segments with constant velocity.  For each path, after certain
intervals, the direction of the particles velocity is changed at
random, with isotropic distribution, corresponding to collision of
randomly distributed pointlike impurities.  The distance of free
propagation between collision is determined at random,
Poisson-distributed $P(l)\propto e^{-l/l_e}$, so that the mean-free
path is $l_e$.  On impact with one of the nanowires walls, reflection
occurs in a specular fashion, by reversing the velocity component
perpendicular to the wall.  The resulting ensemble will consist of
paths which are open (start and end point do not coincide).

\subsection{Evaluation of $\mathcal{M}_B$, $\mathcal{M}_\text{so}$}
After obtaining an ensemble of Boltzmannian paths, for each path the
integrals Eq.~\eqref{eq:modB} or Eq.~\eqref{eq:modSOI} are
evaluated. Because the paths consist of straight line segments, the
evaluation is elementary for each segment, and the integrals
$\mathcal{M}_B$, $\mathcal{M}_\text{so}$ are the products of these
segments. For $\mathcal{M}_B$, these are the phase factors
$e^{i\phi_n}$ accumulated along each segment, while for
$\mathcal{M}_{so}$ we must multiply unitary two-by-two matrices which
describe the spin dynamics along each segment.  When calculating
$\mathcal{M}$ at the same time as generating the path, only the last
position, velocity and accumulated product of
$\mathcal{M}_{B,\text{so}}(t)$ need to be kept in memory.

\subsubsection{Magnetic field}
To be more specific, for magnetic fields we choose the field to point
along the $y$ direction, and the nanowire to lie along either the $x$
or $y$ direction, so that the magnetic field is either perpendicular
or parallel to the nanowires axis.  In the perpendicular case, the
orientation of the nanowire was either such that the magnetic field
penetrated one of the faces perpendicularly, or such that it was
parallel to one of the faces (the difference being a rotation by 30
degrees). It was established that for the resulting $\tau_B$ there is
no significant difference between these two orientations in the
relevant regime.

When choosing the gauge,
\begin{align}
    A(\bm{r}) &= (Bz,0,0)
    \label{eq:si-landau-gauge}
\end{align}
the generation of open paths is sufficient for the evaluation of
$\mathcal{M}_B(t)$ according to Eq.~\eqref{eq:modB}, because the
average $\llangle \mathcal{M}_B(t) \rrangle$ over open and closed
paths is then identical \cite{suppBeenakker1988}. Since open and closed paths
are equivalent in this situation, we use open paths that are easier to
generate numerically than closed paths. In our simulations, we
chose an ensemble size of $2^{14}$ open paths to for averaging.

\subsubsection{Spin-orbit}

For $\llangle \mathcal{M}_{so} \rrangle$ an evaluation with open paths
is not possible, and we have to average over an ensemble of closed
paths, which is created as described in the following. By creating a
number $N$ of open paths of length $L/2$, we can create a set of
$N(N-1)/2$ statistically independent open paths of length $L$, by
pairwise concatenation of two different paths.  We restrict this much
larger set of paths to those which are almost closed (with start and
end point separated not further than $l_e$), and then insert an
additional line segment that closes these paths. If the concatenated
paths are of sufficient length, we assume that the insertion of this
additional line segment with a slightly different length distribution
than the other line segments does not change the ensemble properties
appreciably.  Because we thus could only use a subset of the generated
paths, we chose an ensemble size of $2^{16}$ open paths in this case.
(The size of the ensemble of closed paths decreases with increasing
$L$).

\subsection{Fitting decay times}

Finally, after having created ensembles of open or closed paths as
described above for a set of different path lengths, which we chose to
be logarithmically spaced, $t_n=(1.1)^n \tau_e$ with $n$ integer and
$1\leq t_n/\tau_e \leq 10^6$, we determined the averages $\langle
M_{B,so}(t) \rangle$ and numerically fitted the exponential decays
according to Eqs.~(5) in the main text, resulting in estimates for the
decay times $\tau_B$ and $\tau_{so}$.

\section{3. Validating the numerics against known results}

\subsection{Square nanowire in magnetic field}

To validate the results of our simulations for $M_B$, we also simulate
other geometries, in which results have been found previously,
numerically or analytically.  First, instead of considering hexagonal
nanowires, we change the shape of the nanowire to be square.  If a
square nanowire is placed in a perpendicular magnetic field and has
specularly reflecting walls, we expect the result to be the same as
for a 2D layer, as treated in \cite{suppBeenakker1988}. This is because
reflections on the walls perpendicular to $B$ do not change the
projection of the path along the direction of $B$, and thus are
ineffective.

We should thus reproduce the result of Ref.~\cite{suppBeenakker1988},
which in the ``clean, weak field'' limit reads
\begin{align}
    \frac{\tau_B}{\tau_e} = & 12.1 \frac{l_m^4}{W^3 l_e}
    \label{eq:bvh-weak-clean-layer}
\end{align}
and should hold for $W\ll l_e$ and $l_m\gg \sqrt{W l_e}$.
In Fig.~\ref{fig:bvh} we show simulation results for both perpendicular and
parallel field for a square nanowire. In perpendicular field, the data agrees
to the analytical results in the regime of its validity (the onset of
cross-over to the diffusive case can be seen). Remarkably, in parallel field,
we also observe a $W^{-3}$ dependence, while for hexagonal geometry, the
dependence on $W^\gamma$ has two different $\gamma$ for the two orientations.

\begin{figure}[h]
    \centering
    \includegraphics[width=4in]{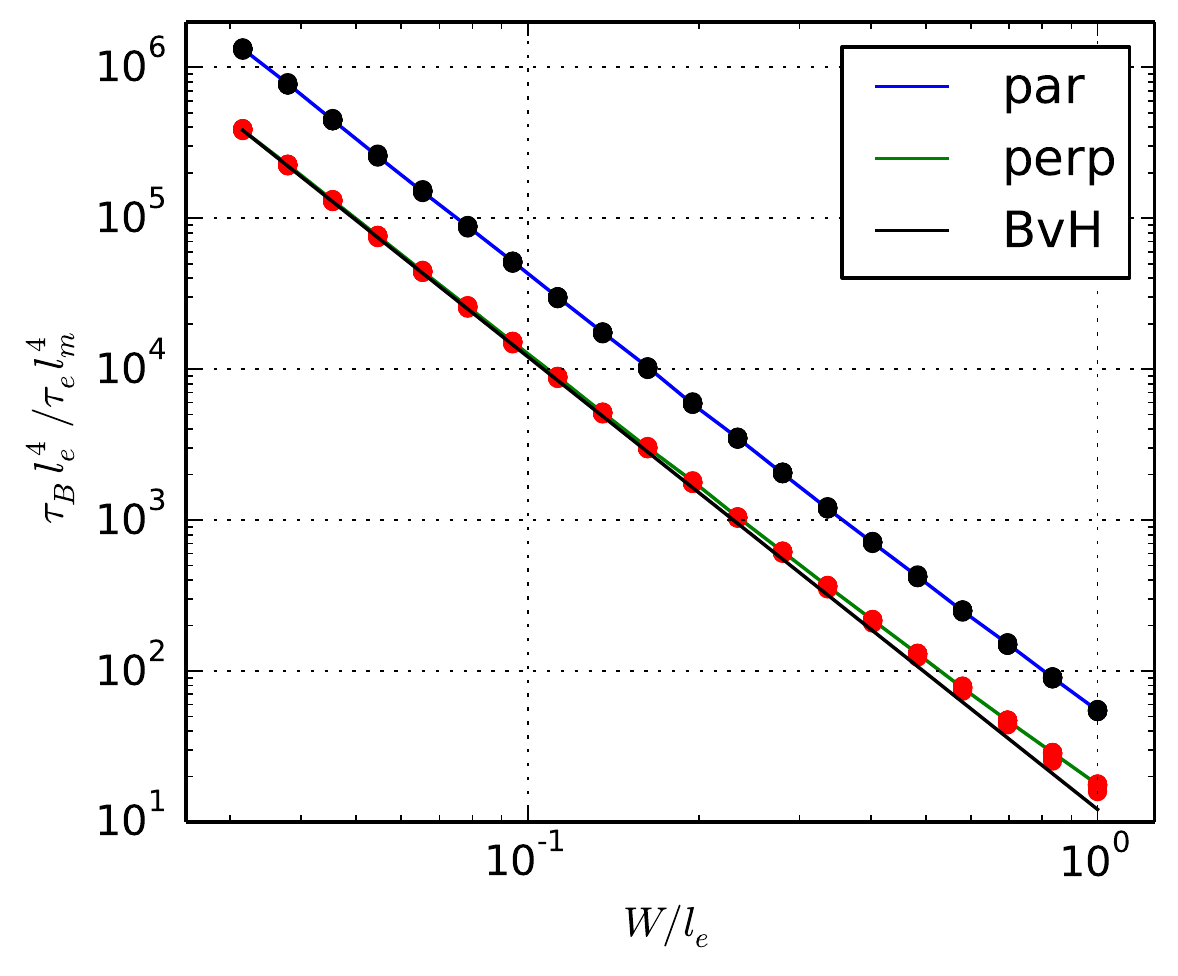}
    \caption{Comparison to the analytical expression by Beenakker and van Houten \cite{suppBeenakker1988}. Data points are shown for different magnetic field $ l_e < l_m  < 10^{1.5} l_e$. The fact that points for different $l_m$ collapse
        shows the expected $l_m^4$ behavior, the solid line is the asymptotic expression Eq.~\eqref{eq:bvh-weak-clean-layer} for $W\ll l_e$. For $W\simeq l_e$, a cross-over to the diffusive regime can be observed.}
    \label{fig:bvh}
\end{figure}

\subsection{Spin-orbit coupling in 2D strip}

To check the calculations of $\mathcal{M}_{so}$, we compare our simulations to
the expression for $\tau_\mathrm{so}$ for two-dimensional diffusive
wires ($l_e \ll W$) with Rashba spin-orbit interaction from Kettemann
\cite{suppKettemann2007}.

When comparing $\tau_\mathrm{so}$ between different sources it is important
to note that different conventions for $\tau_\mathrm{so}$ exist (such as
choosing a factor $4/3$ in Eq.~\eqref{eq:expform}). For
consistency it is thus important to compare physical observables. For
weak antilocalization this is the conductance correction.
In order to describe the case of diffusive wires ($l_e \ll W$) we need
to take the limit $l_e \rightarrow 0$ in Eq.~(1) of the main text.:
\begin{equation}\label{eq:diffusive_us}
\Delta G = - \frac{e^2}{h} \frac{\sqrt{D}}{L} \biggl[
3\, \biggl( \frac{1}{\tau_\varphi} + \frac{4}{3\tau_\text{so}} +
\frac{1}{\tau_B} \biggr)^{-\frac{1}{2}}
\!\!\!- \biggl( \frac{1}{\tau_\varphi} + \frac{1}{\tau_B} \biggr)^{-\frac{1}{2}}
\biggr].
\end{equation}
Kettemann uses a Green's function based approach and arrives at
\cite{suppKettemann2007}:
\begin{equation}\label{eq:diffusive_kettemann}
\Delta G = - \frac{e^2}{h} \frac{\sqrt{D}}{L} \biggl[
2\, \biggl( \frac{1}{\tau_\varphi} +
\frac{1}{2\tau_\text{so}^\text{Ref.~\cite{suppKettemann2007}}} +
\frac{1}{\tau_B} \biggr)^{-\frac{1}{2}}
\!\!\! + \biggl( \frac{1}{\tau_\varphi} +
\frac{1}{\tau_\text{so}^\text{Ref.~\cite{suppKettemann2007}}} +
\frac{1}{\tau_B} \biggr)^{-\frac{1}{2}}
\!\!\!- \biggl( \frac{1}{\tau_\varphi} + \frac{1}{\tau_B} \biggr)^{-\frac{1}{2}}
\biggr].
\end{equation}
In the limit of small spin-orbit splitting, $1/\tau_\mathrm{so} \rightarrow 0$,
both expressions become equal if we identify
\begin{equation}
\tau_\mathrm{so} = 2 \tau_\mathrm{so}^\text{Ref.~\cite{suppKettemann2007}}\,.
\end{equation}
Hence we need to take this factor of 2 into account when comparing our results
to Kettemann's. Taking this factor into account, the expressions
\eqref{eq:diffusive_us} and \eqref{eq:diffusive_kettemann} not only agree
for weak spin-orbit, but also never differ by more than 5\% for all
$\tau_\mathrm{so}$.

Fig.~\ref{fig:kettemann} shows the comparison between the expression given in Ref.~\cite{suppKettemann2007}, which after conversion to the quantities in this paper is
\begin{align}
    \tau_{so}/\tau_{e}&= 3 l_R^4 / W^2,
    \label{eq:kettemann-expressions}
\end{align}
and numerical results we obtained for a diffusive 2D strip for different
spin-orbit strengths.

\begin{figure}
    \centering
    \includegraphics[width=4in]{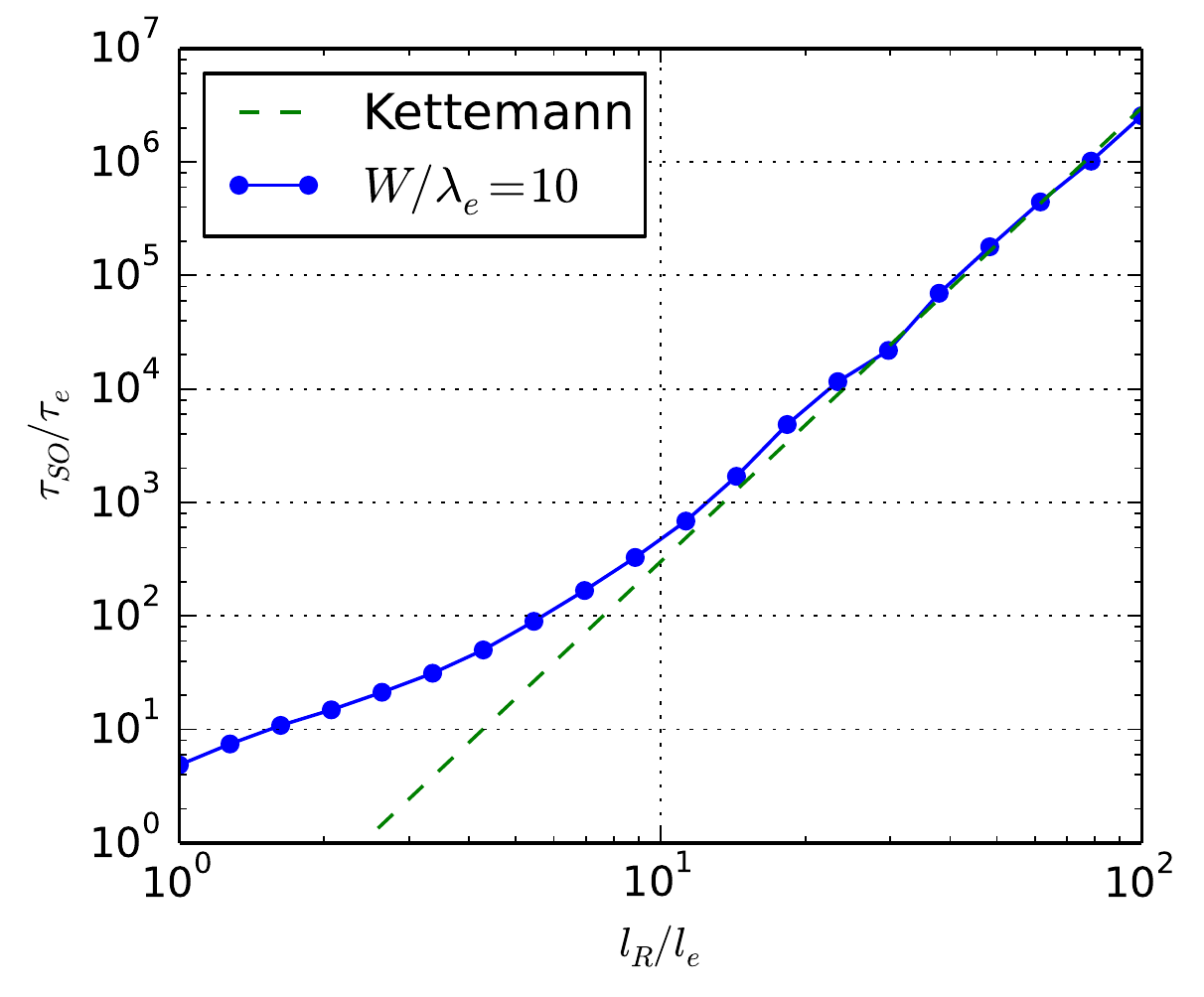}
    \caption{Comparison of the numerical evaluation of $\llangle
      \mathcal{M}_{so}\rrangle$ in a 2D strip (blue dots and line) and
      the diffusive result of Ref.~\cite{suppKettemann2007} (dashed
      line). In the numerics, the width of the strip is $W=10\l_e$, so
      that motion is diffusive.  The Cooperon-based treatment in
      Ref.~\cite{suppKettemann2007} applies for $l_R > W$.}
    \label{fig:kettemann}
\end{figure}

\clearpage

\section{4. Device fabrication and estimations of mobility, mean free path, wire diameter and occupied subbands}

\subsection{Device fabrication}
The nanowire is deposited onto a p$^{++}$-doped Si substrate covered by 285 nm SiO$_2$ (depicted in black in Fig. 3a of the main text). Contacts to the nanowire (green) are made by a lift-off process using electron beam lithography. Contact material is Ti/Au (25/125 nm). After passivation of the nanowire with a diluted ammoniumpolysulfur solution (concentration (NH$_\mathrm{4}$)S$_\mathrm{X}$:H$_\mathrm{2}$O 1:200) the chip is covered with HfO$_2$ (30 nm), deposited by atomic layer deposition. The dielectric is removed at the bonding pads by the writing of an etch mask (PMMA) followed by an HF etch. A top gate (brown) is deposited using a lift-off process with electron beam lithography. Top gate is defined using Ti/Au (25/175 nm). Lastly, an additional layer of Ti/Pt (5/50 nm) is deposited on the bond pads to reduce the chance of leakage to the global back gate. Devices were only imaged optically during device fabrication. SEM imaging was performed only after the measurements.

\subsection{Estimation of mobility, mean free path and $\frac{l_e}{W}$}

Nanowire mobility, $\mu$, is obtained from pinch-off traces using the method described in section 3 of the Supplementary Material of 
\cite{suppPlissard2013}. In short, mobility is obtained from the change of current, or conductance, with gate voltage. We thus extract field-effect mobility, whereby we rely on a fit of the gate trace to an expression for gate-induced transport. This expression includes  a fixed resistance in series with the gated nanowire.

To extract mobility and series resistances from device I (data shown in Fig. 3-5a of the main text and Fig. \ref{WAL_figS1}, Fig. \ref{WAL_figS7}, Fig. \ref{WAL_figS3}, Fig. \ref{WAL_figS4a}, Fig. \ref{WAL_figS4b} of this document) in this way, a gate trace from pinch-off to saturation is needed. However, $I$($V_{BG}$, $V_{TG}$ = 0 V) obtained from Fig. \ref{WAL_figS1}a covers only an intermediate range (see \ref{WAL_figS1}b). Therefore traces at $I$($V_{BG}$, $V_{TG}$ = $-$0.15 V) and $I$($V_{BG}$, $V_{TG}$ = 0.15 V), shown in Fig. \ref{WAL_figS1}b are also used. The three traces then together form a full pinch-off trace (see Fig. \ref{WAL_figS1}c) that is well approximated by Eq. 11 in \cite{suppPlissard2013} for which here an equivalent expression for current $I$ instead of conductance $G$ was used. Here the capacitance between back gate and nanowire $C_{BG}$ = 22 aF, the series resistance $R_S$ = 10 k$\Omega$, the mobility $\mu$ = 12,500 cm$^2$/Vs and the threshold voltage $V_{TG}$ = $-$16.5 V (see Fig. \ref{WAL_figS1}c). Other inputs are source-drain bias $V_{SD}$ = 10 mV and contact spacing $L$ = 2 $\mu$m. The capacitance has been obtained from electrostatic simulations in which the hexagonal shape of the nanowire has been taken into account. The series resistance $R_S$ consists of instrumental resistances (RC-filters and ammeter impedance, together 8 k$\Omega$) and a contact resistance $R_C$. The experimental pinch-off traces are best approximated by $R_C$ = 2 k$\Omega$. Expressions for $I$($V_{BG}$) with $R_C$ = 1 k$\Omega$ and $R_C$ = 3 k$\Omega$, also shown in Fig. \ref{WAL_figS1}c, deviate from the measured pinch-off traces. 

Mobility is also estimated from a linear fit to the top gate pinch-off trace shown in Fig. \ref{WAL_figS1}d. Prior to this fit instrumental and series resistances have been subtracted. From the fit $\mu\sim$ 9,000 cm$^2$/Vs is obtained, using $C_{TG}$ = 1440 aF, obtained from electrostic simulations, and $L$ = 2 $\mu$m. 

Similarly, mobility in device III (see Fig. \ref{WAL_figS1}e, magnetoconductance data shown in Fig. \ref{WAL_figS5} of this supplementary document) is extracted from a fit to the top gate pinch-off trace, giving $\mu\sim$ 10,000 cm$^2$/Vs using $C_{TG}$ = 1660 aF and $L$ = 2.3 $\mu$m. These mobilities are similar to those obtained in InSb nanowires that are gated using only a global back gate \cite{suppPlissard2013}.     

\begin{figure}[h!]
	\centering
	\includegraphics[scale=0.75]{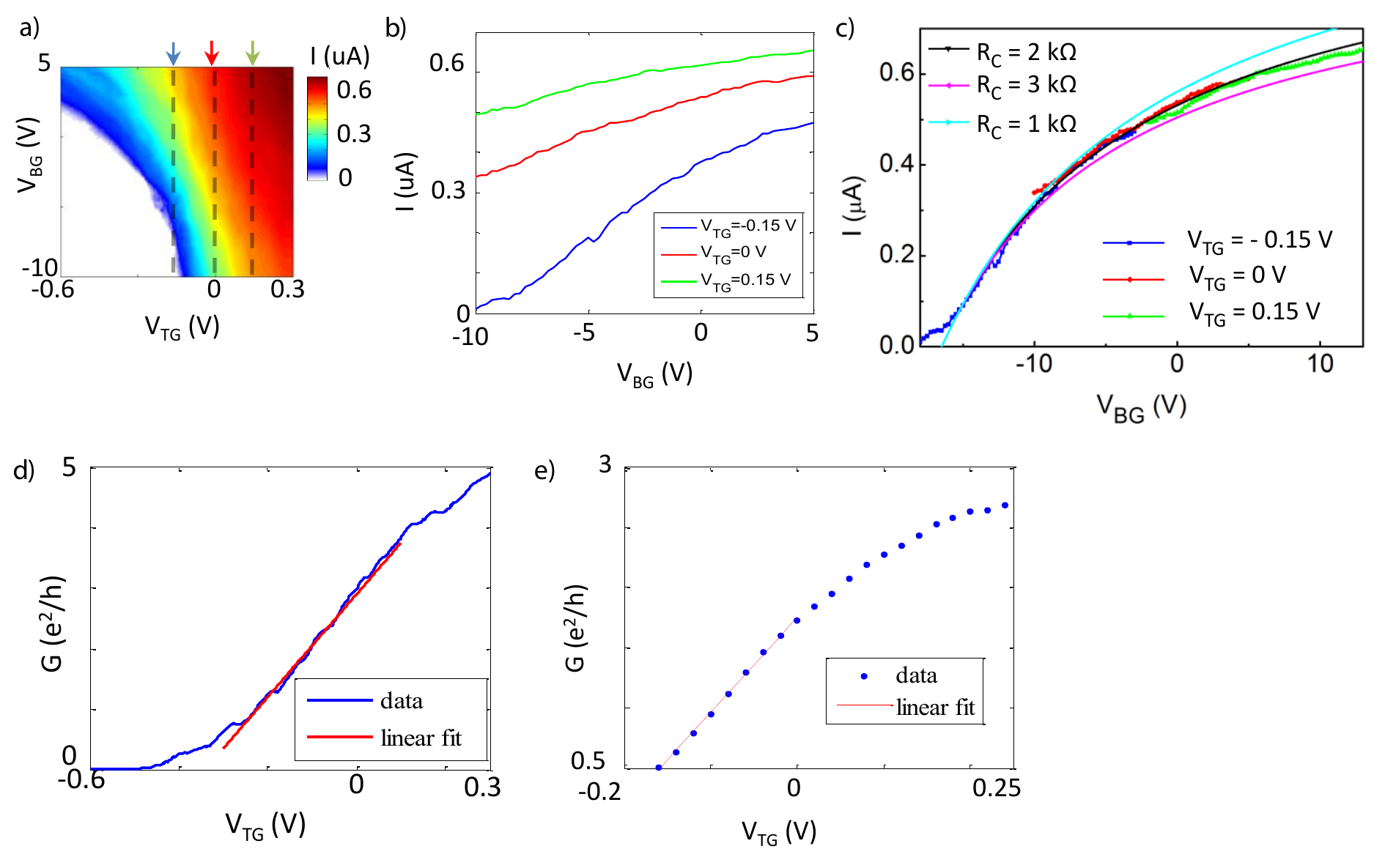}
	\caption{\textbf{a)} Current, $I$, in device I as a function of top gate voltage, $V_{TG}$, and back gate voltage, $V_{BG}$. Cross sections corresponding to the $I$($V_{BG}$) traces in panel b are indicated with arrows. Data taken with source-drain voltage $V_{SD}$ = 10 mV. \textbf{b)} $I$($V_{BG}$) at $V_{TG}$ = 0.15 V, $V_{TG}$ = 0 V and $V_{TG}$ = $-$0.15 V. \textbf{c)} Traces at $I$($V_{BG}$, $V_{TG}$ = $-$0.15 V) (blue) and $I$($V_{BG}$, $V_{TG}$=0.15 V) (green) are displaced by $\Delta V_{BG}$ = $-$8 and  $\Delta V_{BG}$ = 8 V, respectively, chosen such that their current is similar to that of the $I$($V_{BG}$, $V_{TG}$=0 V) trace (red). Data is well approximated by $I(V_{BG})$ (see text) with mobility $\mu \sim$ 12,500 cm$^2$/Vs and contact resistance $R_C$ = 2 k$\Omega$ (black). Traces with larger (3 k$\Omega$, pink) or smaller (1 k$\Omega$, cyan) contact resistance are also shown. \textbf{d)} $G$($V_{TG}$) in device I with $V_{BG}$ = 0 V (blue). A linear fit of the pinch-off traces (red) gives a slope $\frac{dG}{dV_{TG}}$ = 8.5 (e$^2$/h)/V. \textbf{e)} $G$($V_{TG}$) in device III with $V_{BG}$=0 V. A linear fit of the pinch-off traces (red) gives a slope $\frac{dG}{dV_{TG}}$ = 7.9 (e$^2$/h)/V.}
	\label{WAL_figS1}
\end{figure}

Mean free path, $l_e$, is estimated as $l_e$ = $v_F\tau_e$, with $v_F$ the Fermi velocity and $\tau_e$ the scattering time. $\tau_e$ = $\frac{\mu m^*}{e}$, with $e$ electron charge and $m^*$ the effective electron mass in InSb. Assuming a 3D density of states $v_F$=$\frac{\hbar}{m^*}(3\pi^2n)^{\frac{1}{3}}$ with $\hbar$ the reduced Planck constant and $n$ electron density, $n$ is estimated from pinch off traces using $n$ = $\frac{C(V_G-V_{TH})}{eAL}$ with $A$ the nanowire cross section, $V_G$ top or back gate voltage and $V_{TH}$ the threshold (pinch-off) voltage. In this way in device I  $n$ up to $\sim$4$\cdot$10$^{17}$ cm$^{-3}$ are obtained, giving $l_e$ up to $\sim$ 160 nm. This estimate of $n$ agrees reasonably with densities obtained from a Schr\"{o}dinger-Poisson solver (see 'Estimation of the number of occupied subbands'). In device III $n$ up to $\sim$4$\cdot$10$^{17}$ cm$^{-3}$ gives $l_e \sim$ 150 nm. 
Together with the facet-to-facet width $W$ (described in Fig. \ref{WAL_figS2}) these mean free paths yield a ratio $\frac{l_e}{W}$ = 1-2.

\subsection{Nanowire width}

Nanowires were not imaged with scanning electron microscope prior to device fabrication to avoid damage due to electron irradiation. The wire diameter is estimated from a comparison of the nanowire width after fabrication to the nanowire diameter obtained from a number of wires from the same growth batch deposited on a substrate as described in Fig. \ref{WAL_figS2}. 

\begin{figure}[h!]
	\centering
	\includegraphics[scale=0.7]{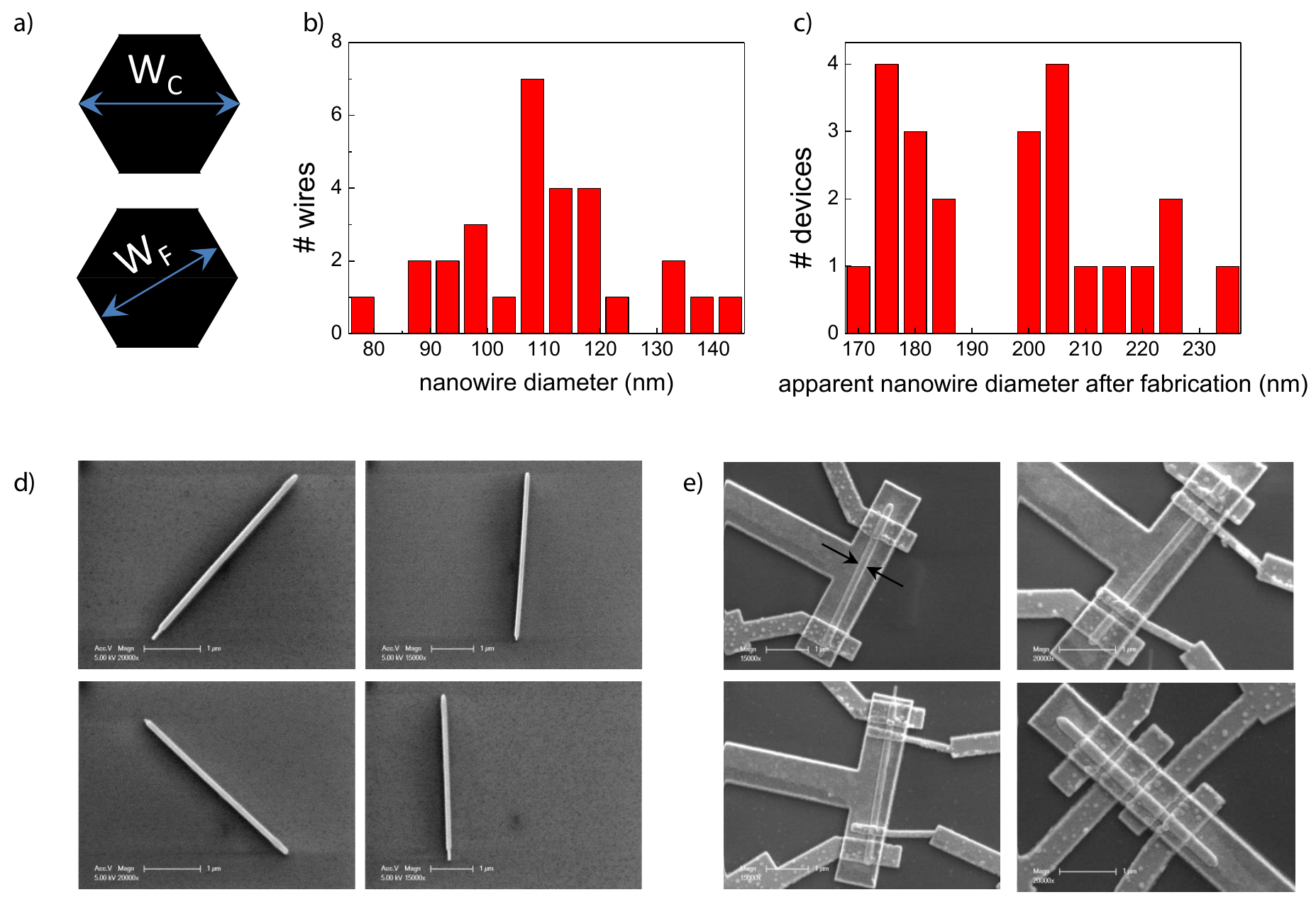}
	\caption{\textbf{a)} Cross-sectional view of hexagonal nanowires with indicated widths $W_C$ and $W_F$. A top view of these nanowires (such as a scanning electron microscope image) shows the width from corner to corner, $W_C$. In our simulations of electron interference in hexagonal nanowires the facet-to-facet width, $W_F$, is used. The two widths are related by $W_F$=$\cos(\frac{\pi}{6})W_C$. \textbf{b)} Distribution of nanowire diameters obtained from scanning electron microscope images of nanowires lying on a substrate. The imaged nanowires are from the same growth batch as the ones used in the experiment. The nanowire diameter is the width of the nanowire when lying on a substrate and thus corresponds to $W_C$ in panel a plus twice the native oxide thickness. Four imaged wires are shown in panel d. Average diameter is 110 nm, standard deviation is 15 nm. \textbf{c)} Distribution of the apparent nanowire diameter after device fabrication. The distribution has been obtained from scanning electron miscroscope images of devices made in the same fabrication run (and thus with the same fabrication recipe) as the ones measured. The apparent diameter increases due to HfO$_2$ and top gate metal deposition.  Average apparent diameter is 197 nm. Device I had an apparent diameter after fabrication of 200 nm, close to the average apparent nanowire device diameter, and therefore its wire diameter is estimated as 110 nm, the average the distribution of wire diameters in panel c. Device III has a diameter after fabrication of 180 nm, which is 17 nm below average. Wire diameter is therefore estimated as 110 $-$ 17 = 93 nm. 
	Wires are covered by a native oxide of $\sim$2.5 nm, giving an InSb diameter $W_C \approx$ 105 nm and $W_C \approx$ 88 nm 
	for device I and device III respectively. Facet-to-facet diameter $W_F$, simply denoted by $W$ throughout the main text, is therefore $W\approx$ 90 nm (device I) and $W_F = W \approx$ 75 nm (device III). The standard deviation of wire diameter  of 15 nm in panel b) is used to define a range of wire diameters, $W\pm$15 nm, for which spin relaxation length, spin precession length and phase coherence length are obtained in Fig. 4 of the main text. \textbf{d)} Scanning electron microscope image of four of the nanowires used to obtain the histogram of nanowire diameters of panel b. \textbf{e)} Scanning electron miscoscope image of four of the devices imaged to obtain the apparent nanowire diameter after fabrication of panel c. The arrows in the upper left image indicate the apparent nanowire diameter.}
	\label{WAL_figS2}
\end{figure}

\clearpage

\subsection{Estimation of the number of occupied subbands}

\begin{figure}
	\centering
	\includegraphics[scale=1]{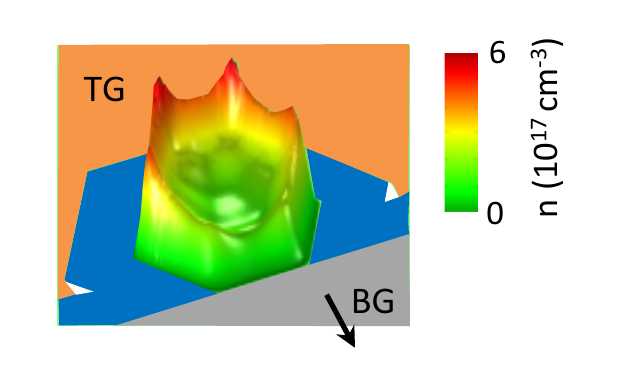}
	\caption{Electron density as a function of the nanowire cross section. Density is obtained from self-consistent Schrodinger-Poisson calculations with $V_{TG}$ = 0.5 V and $V_{BG}$ = 0 V. TG (BG) denotes top (back) gate.}
	\label{WAL_figS6}
\end{figure}

An estimate of the number of occupied subbands is calculated in two ways:

\begin{enumerate}
\item A self-consistent Schrodinger-Poisson calculation yields that 17 subbands contribute to transport at higher device conductance (density profile shown in the inset of Fig. \ref{WAL_figS6}). As contact screening has been neglected in these two-dimensional calculations the actual number of subbands may be slightly lower, but likely several ($\sim$ 10) modes contribute at high device conductance. 
\item The conductance, $G$, of a disordered quantum wire relates to the number of subbands, $N$, as \cite{suppBeenakker1997}

\begin{equation}
G = \frac{NG_0}{1+\frac{L}{l_e}},
\end{equation}

which, using $\frac{L}{l_e} \approx$ 10 -- 20 (obtained from the estimate of $l_e$ above) yields $N \geq$ 25. 
\end{enumerate}

\clearpage

\section{5. Supplementary experimental data}

\subsection{Magnetoconductance traces at constant conductance}

\begin{figure}[h!]
	\centering
	\includegraphics[scale=0.9]{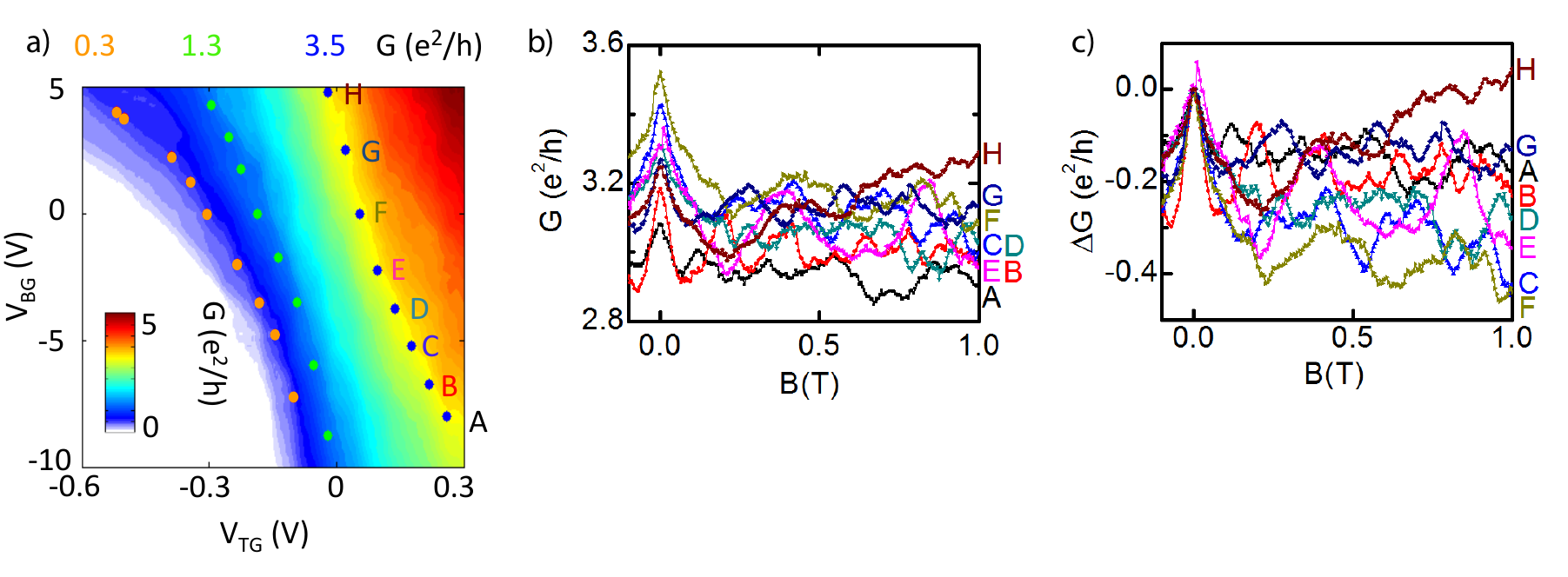}
	\caption{\textbf{a)} Conductance $G$, as a function of top gate
	voltage, $V_{TG}$, and back gate voltage, $V_{BG}$ as shown in Fig. 3(b) of the main text. Dots
	indicate voltages ($V_{BG}$,$V_{TG}$) at which traces in Fig. 4(a) were
	taken (same dot color corresponds to same G). The letters at the dots at $G$ = 3.5 e$^2$/h refer to the magnetoconductance traces shown in panels b) and c). Data obtained with 10 mV voltage bias at a temperature of 4.2 K.\textbf{b)} Magnetoconductance traces taken at the points at $G$ = 3.5 e$^2$/h shown in panel b). Data taken with AC excitation $V_{AC}$ = 100 $\mu$V$_{RMS}$. The difference between the conductance of the dots in panel a) and the conductance of the corresponding magnetoconductance traces in panel b) is likely due to the difference in source-drain bias between both measurements. Also at other conductances (for instance at the green and orange dots in panel a) magnetconductance traces generally show a conductance lower than those obtained in the gate-gate plot of panel a) by a similar amount. For each of these traces the conductance denoted on the vertical axis of Fig. 4a and that on the horizontal axis of Fig. 4b-d is the conductance of the equiconductance points of Fig. 3b of the main text. \textbf{c)} Magnetoconductance traces of panel b) normalized to $\Delta G$($B$ = 0) = 0. By averaging over these traces the blue trace of Fig. 4a of the main text ($G$ = 3.5 e$^2$/h) is obtained.}
	\label{WAL_figS7}
\end{figure}

\subsection{Spin relaxation and phase coherence length obtained from top gate averaging in device I} 

\begin{figure}[h!]
	\centering
	\includegraphics[scale=0.8]{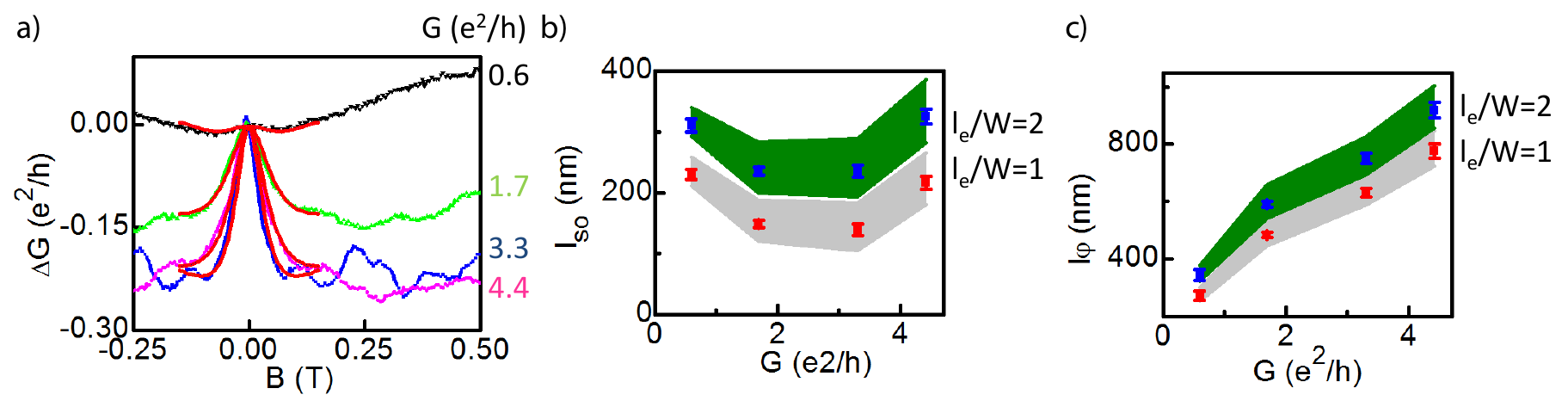}
	\caption{\textbf{a)} Magnetoconductance traces obtained after taking MC traces with top gate voltage spacing $\Delta V_{TG}$ = 20 mV between $V_{TG}$ = 0.34 V and $V_{TG}$ = $-$0.42 V and averaging 9 subsequent traces. $V_{BG}$=0 V. Averaged MC traces have been centered to $\Delta G$=0 at $B$ = 0 T. $G$($B$=0.5 T) is indicated on the right. Red curves are fits to Eq. 1 of the main text wherein Eqs. 2 and 3 of the main text have been used to obtain $l_B$, using $l_e/W = 2$ and $W = 90$nm. \textbf{b)} Spin relaxation length, $l_{so}$, obtained from the fits of panel a) ($\frac{l_e}{W}$ = 2, red points) and obtained from fits with $\frac{l_e}{W}$ = 1 (blue points). Standard deviation of the fit outcomes are indicated. The distribution around the blue and red points (in green and gray, respectively) is given by the spin-orbit lengths obtained from fits with an effective width 15 nm smaller or larger than the expected wire width $W$ = 90 nm. \textbf{c)} Phase coherence length, $l_{\varphi}$, obtained from fits of panel a). Figure formatting (colors, standard deviation and wire diameter dependence) is the same as in panel b).}
	\label{WAL_figS3}
\end{figure}

\subsection{Phase coherence and spin relaxation length at $T$ = 0.4 K}

\begin{table}[h]
\begin{tabular} {| l | l | r |r |}
\hline
G (e$^2$/h) & $\frac{l_e}{W}$ & $l_{so}$ (nm) & $l_{\varphi}$ (nm) \\ \hline
3.9 & 1 & 95 $\pm$ 18 & 1078 $\pm$ 32\\
 & 2 & 205 $\pm$ 16 & 1174 $\pm$ 39\\
2.6 & 1 & 171 $\pm$ 26 & 805 $\pm$ 52 \\
 & 2 & 380 $\pm$ 29 & 937 $\pm$ 60 \\ \hline
\end{tabular}
\caption{Spin relaxation length, $l_{so}$, and phase coherence length, $l_{\varphi}$, obtained from fits to the traces in Fig. 4a of the main text. $\frac{l_e}{W}$ denotes the ratio of mean free path, $l_e$, to wire width, $W$.}
\label{WAL_table}
\end{table}

\clearpage

\subsection{Magnetoconductance in parallel and perpendicular field in device I}

\begin{figure}[h!]
	\centering
	\includegraphics[scale=0.85]{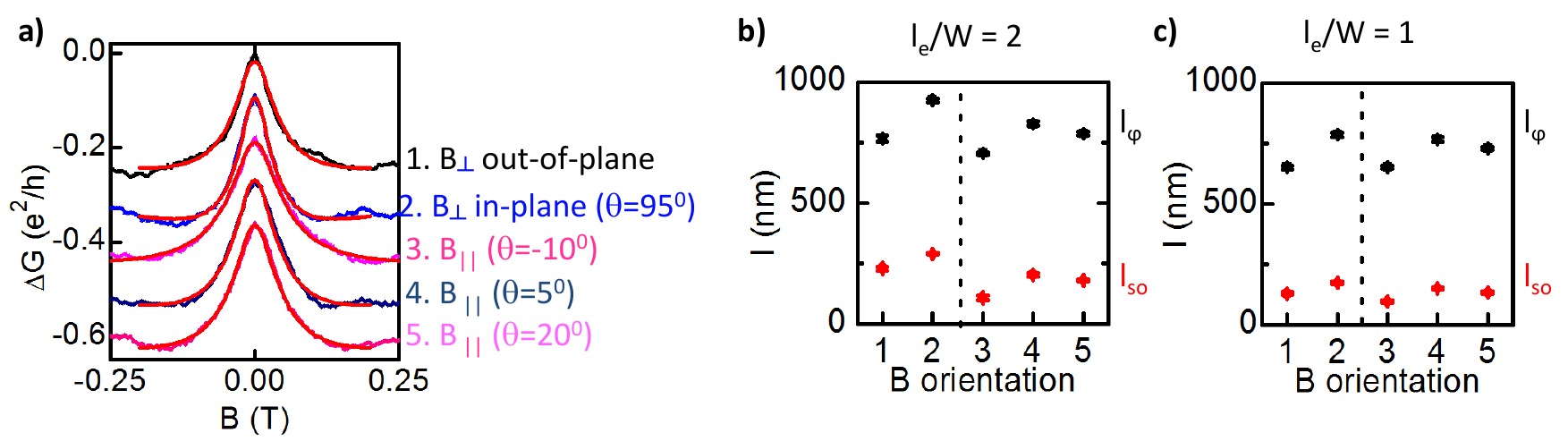}
	\caption{\textbf{a)} MC with parallel and perpendicular magnetic field orientation. Out-of-plane, $\perp$, (in-plane, $\parallel$,) denotes an orientation of the magnetic field (parallel) perpendicular to the substrate plane. $\theta$ denotes the in-plane angle of the magnetic field w.r.t. the nanowire. As the uncertainty in orientation of the in-plane magnetic field is 20$^\circ$ three parallel magnetoconductance traces with $|\theta|\leq$20$^\circ$ are shown. Each MC trace is an average of 7 traces taken at the same conductance $G$=3.5 e$^2$/h by varying top and back gate voltage similar to the MC data of Fig. 2 of the main text. No systematic change of MC along these equiconductance points was observed. As in device II (Fig. 5c of the main text) also here WAL in parallel and perpendicular magnetic field are very similar. Red curves are fits to Eq. 1 of the main text (in which Eqs. 2 and 3 of the main text have been used for $l_B$, with values of $C$ corresponding to parallel or perpendicular magnetic field orientation), using $\frac{l_e}{W}$ = 1 and $W$ = 90 nm. \textbf{b)} Spin relaxation length (red) and phase coherence length (black) obtained from fits of the MC traces in panel a using $\frac{l_e}{W}$ = 2. B orientation numbers correspond to the traces numbered 1 to 5 in panel a. \textbf{c)} Spin-orbit length (red) and phase coherence length (black) obtained from fits of the MC traces in a) using $\frac{l_e}{W}$ = 1. The slightly wider WAL peak in parallel magnetic field yields better agreement with $\frac{l_e}{W}$ = 1 as spin-orbit lengths and phase coherence lengths obtained in parallel and perpendicular field with $\frac{l_e}{W}$ = 1 are more similar than when assuming $\frac{l_e}{W}$ = 2.}
	\label{WAL_figS4a}
\end{figure}

\begin{figure}
	\centering
	\includegraphics[scale=0.90]{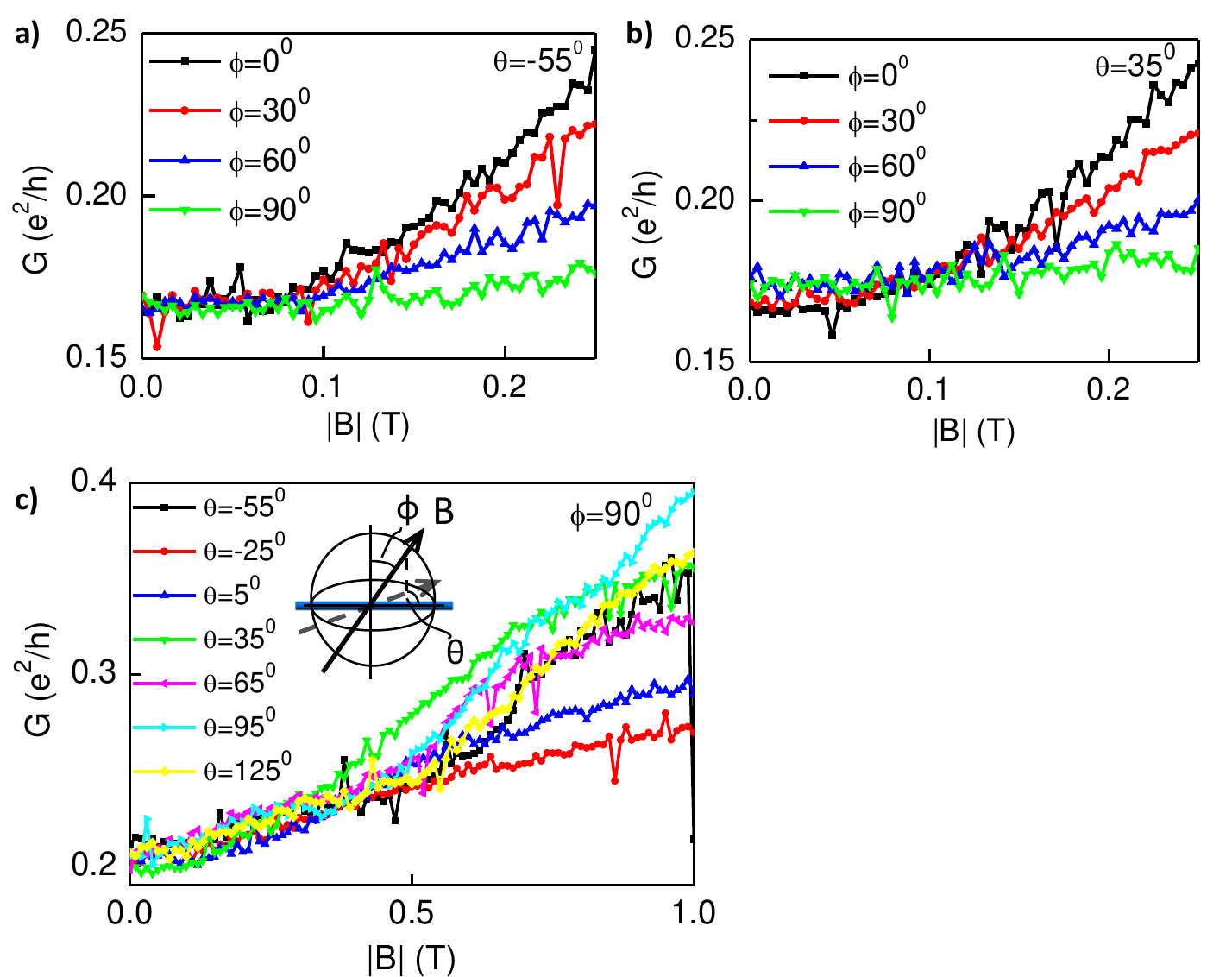}
	\caption{\textbf{a)} MC as a function of out-of-plane angle, $\phi$, with in-plane angle w.r.t. the nanowire $\theta$ = $-$55$\pm$20$^{\circ}$.  Angles $\theta$ and $\phi$ are shown in the schematic drawing in the inset of panel c. Out-of-plane (in-plane) denotes an orientation of the magnetic field (parallel) perpendicular to the substrate plane. $\phi$ = 0$^\circ$ (90$^\circ$) is magnetic field perpendicular to (parallel to) the substrate plane. \textbf{b)} MC as a function of out-of-plane angle $\phi$ with in-plane angle w.r.t. nanowire $\theta$=35$\pm$20$^\circ$. While weak anti-localization is (nearly) independent of magnetic field orientation, here we find that the suppression of weak localization by the magnetic field becomes less effective when rotating the field from perpendicular to parallel to the substrate plane. \textbf{c)} MC as a function of in-plane angle $\theta$. Although the suppression of weak localization by magnetic field is much less effective for all magnetic fields oriented parallel to the substrate plane, a closer inspection shows that the magnetic field dependence is weakest when the magnetic field is approximately aligned with the nanowire. We suggest that the difference in dependence on magnetic field orientation between WAL and WL is due to a difference in charge distribution: while at the larger device conductance at which weak anti-localization is observed many subbands all across the nanowire cross section contribute to transport (see the inset of Fig. 2d of the main text), at low conductance, when weak localization is seen, transport takes place only a few modes, confined to a small region of the nanowire cross section. The low conductance situation may resemble a two-dimensional system, in which only the magnetic field component perpendicular to the substrate leads to a suppression of WL. This would lead to the reduction of positive MC when rotating the magnetic field from out-of-plane to in-plane. In all panels $V_{TG}$ = $-$0.36 V, $V_{BG}$ = 0 V. The difference in $G$($B$ = 0 T) between panels a-b and c is due to a slight device instability at low conductance or due to hysteresis when sweeping $V_{TG}$.}
	\label{WAL_figS4b}
\end{figure}

\clearpage

\subsection{Device III: reproducibility of extracted spin relaxation and phase coherence lengths}

\begin{figure}[h!]
	\centering
	\includegraphics[scale=0.75]{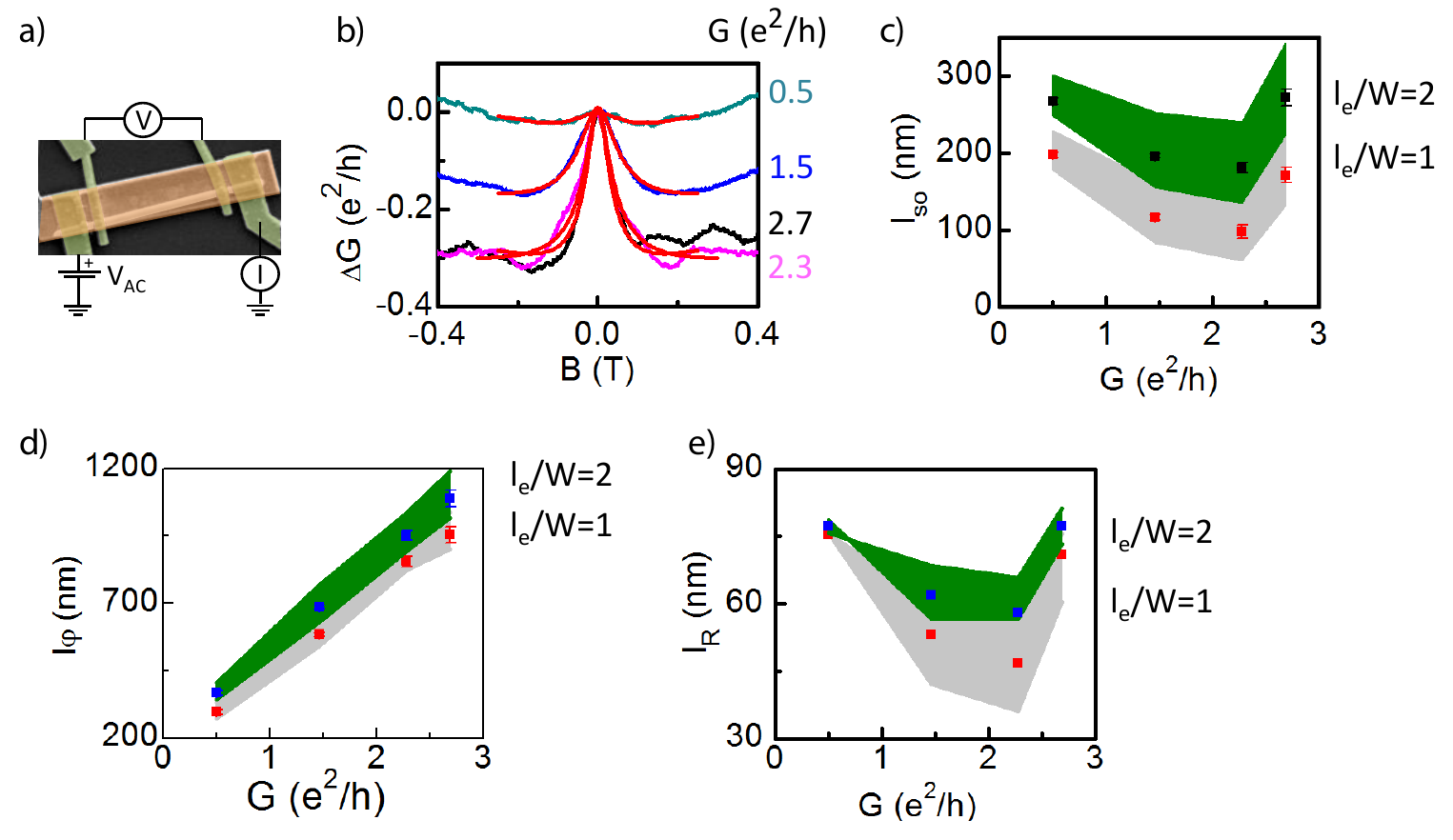}
	\caption{\textbf{a)} False colour scanning electron microscope image of device III. A voltage bias, $V_{AC}$, is applied across the outer contacts, after which simultaneously the current, $I$, through the device and the voltage across the inner contacts, $V$, is measured. Subsequently conductance $G$=$\frac{I}{V}$ is determined. \textbf{b)} Averaged MC traces obtained after taking MC traces with top gate voltage spacing $\Delta V_{TG}$ = 20 mV between $V_{TG}$ = 0.3 V and $V_{TG}$ = $-$0.22 V and averaging 7 subsequent traces. $V_{BG}$=0 V. $G$($|B|$ = 0.5 T) is indicated. Red curves are fits to Eq. 1 of the main text wherein Eqs.2 and 3 of the main text have been used to obtain $l_B$, using $l_e/W = 1$ and $W = 75$nm. \textbf{c)} Spin relaxation length, $l_{so}$, obtained from the fits of panel b) ($\frac{le}{W}$ = 1, blue points) and obtained from fits with $\frac{le}{W}$ = 2 (red points). Standard deviation of the fit outcomes is indicated. The distribution around the blue and red points (in green and gray, respectively) is given by the spin-orbit lengths obtained from fits with an effective width 15 nm smaller or larger than the expected wire width $W$ = 75 nm. \textbf{d)} Phase coherence length, $l_{\varphi}$, obtained from the fits of panel b) ($\frac{le}{W}$ = 1, blue points) and obtained from fits with $\frac{le}{W}$ = 2 (red points). Figure formatting is the same as in panel c. \textbf{e)} Spin precession length, $l_R$, as a function of device conductance, $G$, extracted from the spin relaxation lengths of panel c. Figure formatting is the same as in panel c. When assuming $W$ = 90 nm the $\frac{\tau_{so}}{\tau_e}$ corresponding to the $l_{so}$ at $G$ = 2.3 e$^2$/h are below the simulation range. In this case the $l_R$ corresponding to the lowest simulated value of $\frac{\tau_{so}}{\tau_e}$ has been chosen.}
	\label{WAL_figS5}
\end{figure}

\clearpage

\section{6. Topological gap as a function of mobilty and spin-orbit strength}

We follow the theoretical analysis of Ref.~\cite{suppSau2012} to compute
the maximum topological gap that can be achieved at a given mobilty
$\mu$ and spin-orbit strength $\alpha_\text{R}$. One should only be
careful to note that the definition of $E_\text{SO}$ in \cite{suppSau2012}
differs by a factor of 4 from ours. Whenever we refer to $E_\text{SO}$ here,
we use our definition from the main text.

\begin{figure}[h]
\includegraphics[width=0.9\linewidth]{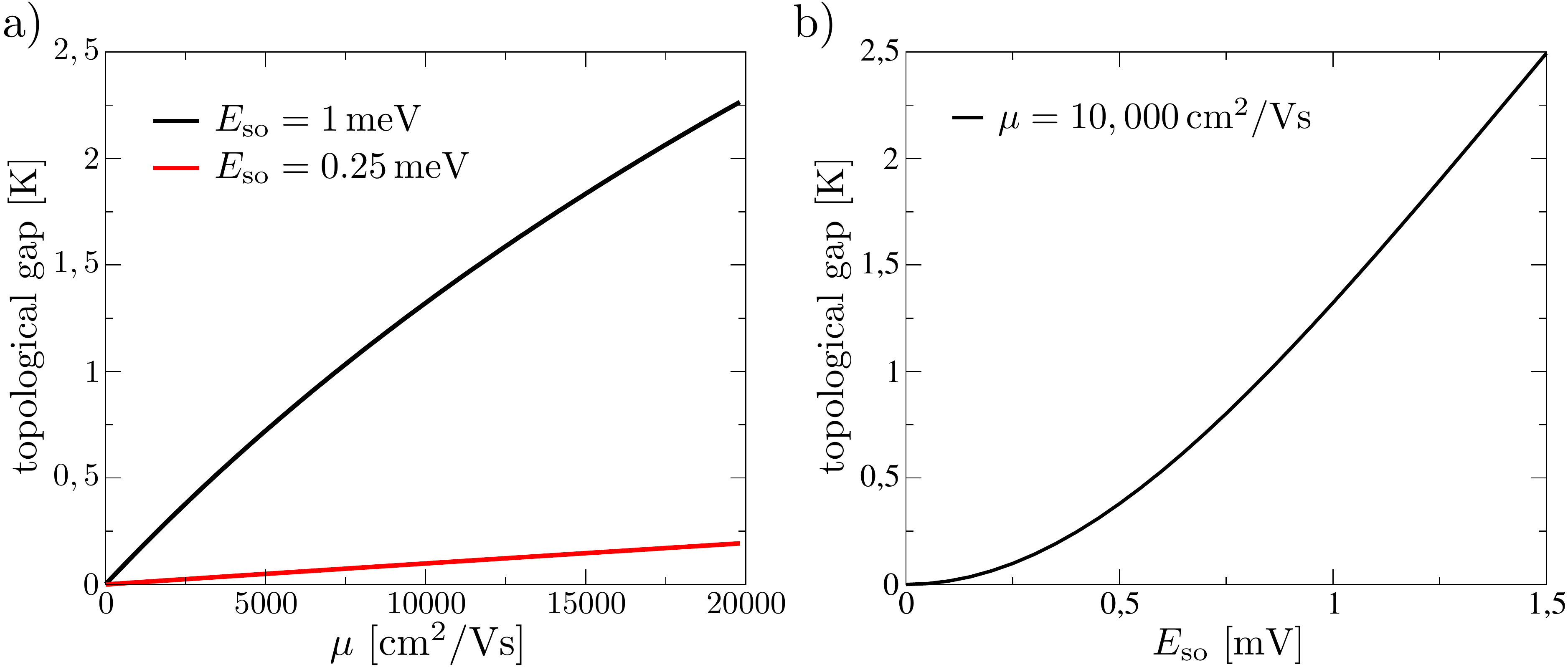}
\caption{\textbf{a)} Topological gap as a function of mobility for different
  values of $E_\mathrm{so}$. \textbf{b)} Topological gap as a function of
  $E_\text{so}$ for a fixed mobility of
  $\text{10,000}\,\text{cm$^2$/Vs}$.  The remaining parameters were
  chosen to be suitable for InSb nanowires in proximity to NbTiN:
  effective mass $m^*=0.014 m_e$ and superconducting gap
  $\Delta=30\,\text{K}$.}\label{fig:saupaper}
\end{figure}

In Fig.~\ref{fig:saupaper}a we show the topological gap as a
function of mobility for the spin-orbit energies estimated in the main
text, with parameters suitable for the Majorana experiments in
Ref.~\cite{suppMourik2012}.  We observe a nearly linear dependence of the
topological gap on mobility for these parameters. The topological gap
can be rather sizable, and we find gaps of order $1\,\text{K}$ for a
moderate mobility of $\mu = \text{10,000}\,\text{cm$^2$/Vs}$ for
$E_\text{so}=1\,\text{meV}$. From the figure it is also apparent that
the topological gap depends rather strongly on $E_\text{so}$.

We investigate the $E_\text{so}$-dependence of the topological gap in
Fig.~\ref{fig:saupaper}b. At a mobility of
$\text{10,000}\,\text{cm$^2$/Vs}$ the topological gap depends roughly
quadratically on $E_\text{so}$ up to $E_\text{so} \sim 1\,\text{meV}$,
i.e.~the topological gap increases as $\alpha_R^4$. This is in stark
contrast to the clean case where the topological gap depends linearly
on $\alpha_R$.

The different dependences of the topological gap on mobility (linear)
and spin-orbit strength (to the fourth power) indicates that for
current devices it may be more efficient to attempt to improve
spin-orbit strength rather than mobility.


\begin{thebibliography}{50}
\bibitem{footnote1} Exceptions are the smallest values of $l_\mathrm{so}$ at
  $G = 2.4$ and $2.8 e^2/h$: When assuming a wire width larger than
  the expected value ($W = 105\,\text{nm}$) we find
  $\frac{\tau_\mathrm{so}}{\tau_e} \sim 1$. In this case the $l_\mathrm{R}$
  corresponding to the lowest simulated value of
  $\frac{\tau_\mathrm{so}}{\tau_e}$ have been chosen as a lower bound.
\end{thebibliography}
\end{document}